\documentclass[traditabstract]{aa}

\usepackage{graphicx}  
\usepackage{dcolumn}   
\usepackage{bm}        
\usepackage{amssymb}   
\usepackage[utf8]{inputenc}
\usepackage{amsmath}
\usepackage{amssymb}
\usepackage{color}
\usepackage{rotating}
\usepackage{float}
\usepackage{wrapfig}
\usepackage{enumitem}
\usepackage{listings}
\usepackage{caption}
\usepackage{wrapfig}
\usepackage{array}
\usepackage{wasysym}
\usepackage{braket}
\usepackage[titletoc]{appendix}
\usepackage{tikz}
\usepackage{mathtools}
\usetikzlibrary{calc}
\usepackage{makeidx}
\usepackage{slashed}

\usepackage{subcaption}
\usepackage[round]{natbib}
\begin{document}

\title{A first attempt to differentiate between modified gravity and modified inertia with galaxy rotation curves}

\author{Jonas Petersen
	\inst{1}\fnmsep\thanks{petersen@cp3.sdu.dk}
	\and
	Federico Lelli \inst{2,3}\fnmsep\thanks{LelliF@cardiff.ac.uk}
}

\titlerunning{A first attempt to differentiate between modified gravity and modified inertia}
\authorrunning{J. Petersen\inst{\ref{i1}} and F. Lelli}

\institute{$CP^3$-Origins, University of Southern Denmark, Campusvej 55, DK-5230 Odense M, Denmark\\
	\and
	{European Southern Observatory, Karl-Schwarschild-Strasse 2, 85748 Garching bei Munchen, German\\
		\and
		School of Physics and Astronomy, Cardiff University, Queens Buildings, The Parade, Cardiff, CF24 3AA, UK\\}
}

\date{Received September 15, 2019; accepted 19 December, 2019}

% \abstract{}{}{}{}{} 
% 5 {} token are mandatory

\abstract{The phenomenology of modified Newtonian dynamics (MOND) on galaxy scales may point to more fundamental theories of either modified gravity (MG) or modified inertia (MI). In this paper, we test the applicability of the global deep-MOND parameter $Q$ which is predicted to vary at the $10\%$ level between MG and MI theories. Using mock-observed analytical models of disk galaxies, we investigate several observational uncertainties, establish a set of quality requirements for actual galaxies, and derive systematic corrections in the determination of $Q$. Implementing our quality requirements to the SPARC database yields $15$ galaxies, which are close enough to the deep-MOND regime as well as having rotation curves that are sufficiently extended and sampled. For these galaxies, the average and median values of $Q$ seem to favor MG theories, albeit both MG and MI predictions are in agreement with the data within $1.5\sigma$. Improved precision in the determination of $Q$ can be obtained by measuring extended and finely-sampled rotation curves for a significant sample of extremely low-surface-brightness galaxies.}


\maketitle

\section{Introduction}
The missing mass problem is established by a host of astronomical observations, including the dynamical behavior of galaxy clusters~\citep{Zwicky:1933gu,Clowe:2006eq}, the rotation curves of disk galaxies~\citep{Rubin1978, Bosma,Albada}, and the properties of the cosmic microwave background~\citep{Ade:2015xua,Komatsu:2003fd}. Solutions to the missing mass problem have been proposed in the form of either some unobserved mass (dark matter) or a modification to the classical laws of dynamics. The former option has been vastly explored and led to the standard cosmological model $\Lambda$ cold dark matter ($\Lambda$CDM), which provides a good description of the Universe on large scales but has some problems on galaxy scales~\citep{Bullock:2017xww}. The latter option led to the development of modified Newtonian dynamics (MOND;\citealt{Milgrom:1983ca}), in which the classical laws of dynamics are modified at small accelerations (see e.g., \citealt{Sanders:2002pf,Famaey:2011kh,Milgrom:2014} for reviews of MOND). MOND is primarily motivated by the close relationship between the baryonic matter and the observed dynamical behavior of galaxies~\citep{Tully:1977fu,Faber1976,McGaugh:2000sr,McGaugh:2016leg,Lelli:2017vgz} as well as the existence of a characteristic acceleration scale ($a_0\sim 1.2\cdot 10^{-10}\frac{m}{s^2}$) below which the dark matter effect appears~\citep{Sanders1990,McGaugh:2004aw,McGaugh:2016leg}.\newline 
The MOND phenomenology can be interpreted either as a modification of inertia (MI) or a modification of gravity (MG). In the context of the nonrelativistic action of the theory, MI and MG differ in that the former alters the kinetic term, whereas the latter alters the potential term~\citep{Milgrom:2002tu,Milgrom:2005mc}. The first full-fledged MG theory was proposed by \citet{Bekenstein:1984tv} and it is commonly known as AQUAL due to the aquadratic form of the Lagrangian. Another MG theory was proposed by \citet{Milgrom:2009gv} and is dubbed quasi-linear MOND (QMOND). On the other hand MI theories has proven to be more difficult to build. \citet{Milgrom:1992hr}, however, demonstrated that MI theories must be strongly nonlocal and that for purely circular orbits, MI leads to the algebraic MOND equation
\begin{equation}
g_{obs}\mu\bigg(\frac{g_{obs}}{a_0}\bigg)=g_N,
\label{eq1}
\end{equation}
where $g_{obs}$ is the observed kinematic acceleration, $g_N$ is the classical Newtonian gravitational acceleration due to the baryons, $\mu$ is an arbitrary interpolation function that reproduces the Newtonian regime at large accelerations, that is $\mu\rightarrow 1$ for $g_{obs}/a_0\rightarrow \infty$, and asymptotic flat rotation curves at low accelerations, that is $\mu\rightarrow\frac{g_{obs}}{a_0}$ for $g_{obs}/a_0\rightarrow 0$. In the MG context, instead, Equation \eqref{eq1} is only valid for highly symmetric mass distributions, such as in the case of spherical geometry~\citep{Bekenstein:1984tv}.\newline 
Both MI and MG have struggled on larger scales, for example in accounting for the entire gravitational anomaly in galaxy clusters~\citep{Sanders:2002ue} and cosmological observations~\citep{Skordis:2005xk,Dodelson:2006zt,Dodelson:2011qv}. However, the two theories remain a popular source of inspiration for model building (e.g., see~\citealt{Chashchina:2016wle,Edmonds:2017zhg,Dai:2017unr,Cai:2017buj,Berezhiani:2017tth,Costa:2019pbz}) due to their ability to provide an intuitive explanation for galactic data. Recently MI has been debated in regards to galactic scales, with \citet{McGaugh:2016leg} and \citet{Li:2018tdo}, presenting evidence in favor of MI and \citet{Petersen:2017klw}, and \citet{Frandsen:2018ftj}, \citet{Petersen:2019obe} presenting evidence in opposition with MI.\newline\newline

\noindent In principle, one could distinguish between MG and MI by performing detailed rotation-curve fits under the two different prescriptions and compare the corresponding residuals. This approach, however, relies on the knowledge of the galaxy distance and disk inclination, which often dominate the error budget (e.g., \citealt{Lelli:2017vgz}, \citealt{Li:2018tdo}) and may well exceed the expected difference of $\sim10\%$ expected between MG and MI~\citep{Brada:1994pk}.\newline

\noindent In this paper, we investigate the possibility of differentiating between MI and MG theories using a parameter proposed by~\citet{Milgrom:2012rk}
\begin{equation}
Q\equiv \frac{2\pi}{m_{bar}v_\infty^2}\int_{0}^{\infty} r\Sigma_{bar}(r) v_{obs}^2(r)dr,
\label{q1}
\end{equation} 
where $v_{obs}$ is the observed rotation velocity, $\Sigma_{bar}$ is the baryonic mass surface density (gas plus stars), $m_{bar}$ is the total baryonic mass, and $v_\infty$ is the asymptotically flat velocity obtained at $r\rightarrow \infty$. \citet{Milgrom:2012rk} shows that for disk-only galaxies everywhere in the deep MONDian limit (DML, maximum acceleration $\ll a_0$), MI predicts $Q^{(MI)}\approx 0.73\pm 0.01$ whereas MG predicts $Q^{(MG)}=\frac{2}{3}$. Thus, there is potential for differentiating between the two model classes by measuring $Q$ in low density, low acceleration galaxies. The result $Q^{(MG)}=2/3$ has general validity and relies only on the form of the deep-MOND virial relation, which is the same for MG theories like AQUAL and QMOND~\citep{Zhao:2010dw}. The result for MI, instead, is not general but depend on the adopted mass distribution. \Citet{Milgrom:2012rk}, however, finds that $Q^{(MI)}$ is very close to $0.73$ for a large set of realistic mass density profiles for disk galaxies. A major advantage of the $Q$ parameter is that uncertainties due to galaxy distance and disk inclination cancel out due to the normalization factors $m_{bar}$ and $v_{\infty}$. The drawback, however, is that both $v_\infty$ and the integral are defined for $r\rightarrow \infty$. Thus, one needs to quantify the systematic effects introduced by measuring $Q$ in finite-size galaxies with finite spatial resolution.

\noindent In section \ref{app:mock}, we create a set of mock galaxies to quantify systematic effects in the estimate of $Q$. In section \ref{sec:Qmes}, we present a first attempt at measuring $Q$ for galaxies in the SPARC database. Finally, in section \ref{sec:discussion}, we draw our conclusions and discuss future prospects.

\section{Mock galaxies}
\label{app:mock}
\begin{figure}[H]
	\centering
	\captionsetup{width=0.5\textwidth}
	\includegraphics[width=0.48\textwidth]{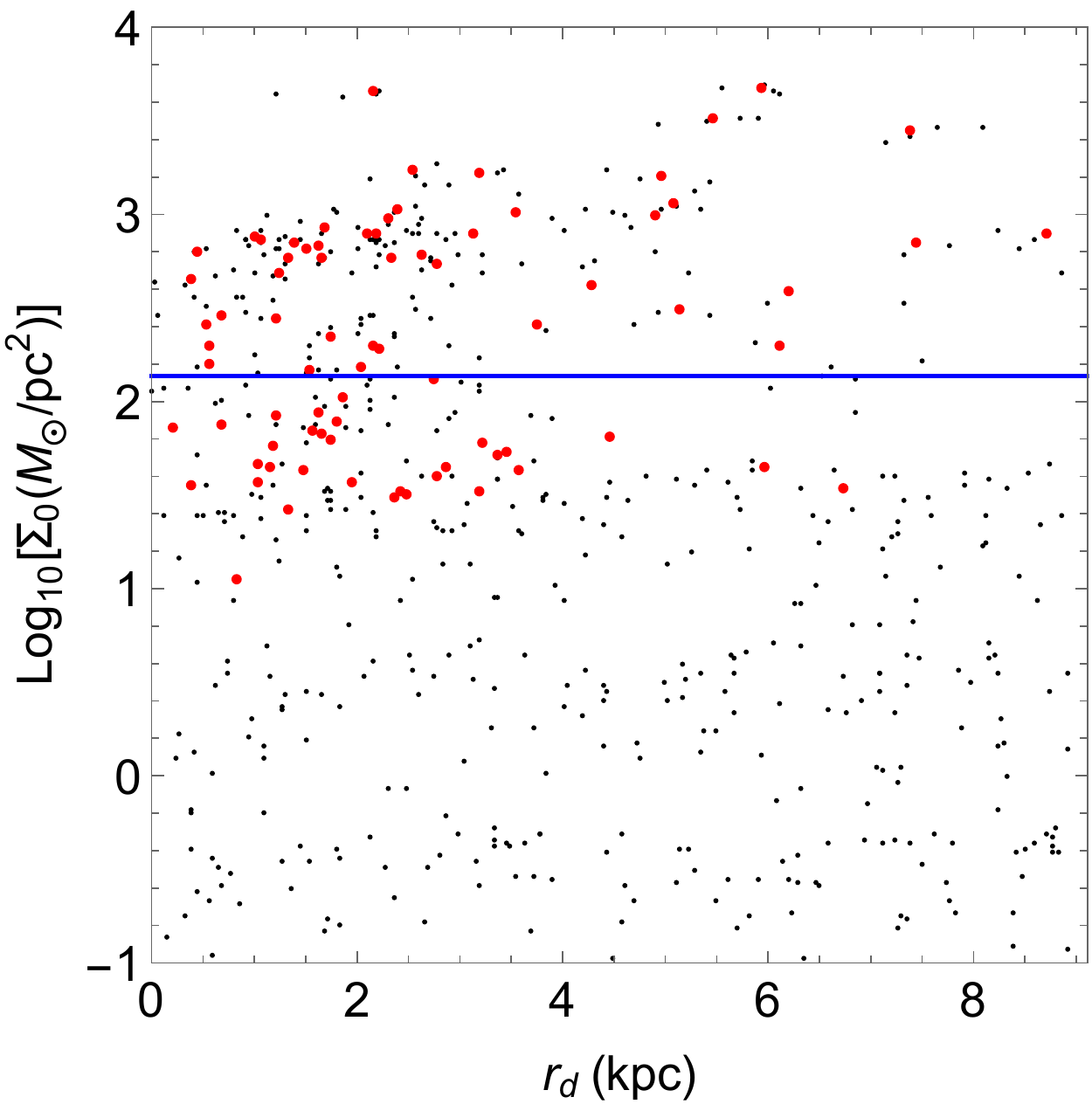}
	\caption{Distribution of real (red) and mock (black) galaxies in the $r_d-\Sigma_d$ plane. The blue line denotes the scale of MOND in terms of the characteristic surface density $\Sigma_M=\frac{a_0}{2\pi G_N}$.}
	\label{fig:1}
\end{figure}

In the following, we create a set of mock galaxies in order to quantify a number of systematic uncertainties in the derivation of $Q$: (1) The effect of deviating from the theoretical deep MOND limit (DML), (2) the effect of the finite-size of galaxy disks and (3) the effect of the finite spatial resolution. We will show that these uncertainties can be accounted for by using the following equation:
\begin{equation}
Q^{(c)}\approx Q^{(m)}-\Delta Q^{(DML)}+\Delta Q^{(range)}-\Delta Q^{(res.)},
\label{qc}
\end{equation}
where $Q^{(m)}$ and $Q^{(c)}$ are the measured and corrected $Q$ values, respectively. $\Delta Q^{(DML)}$, $\Delta Q^{(range)}$ and $\Delta Q^{(res.)}$ are correction factors that can be applied as long as $\max[g_{obs}]\lesssim 0.4a_0$ (acceleration scale requirement), $\max[r]\gtrsim3r_d$ (sampled range requirement) and a spacing between sampled points smaller than $1.3r_d$ (resolution requirement). \newline
For the mock data, let the gas and stellar mass densities be represented by a single disk profile characterized by a central disk mass density and a disk scale length. The mock data are generated by adding Gaussian noise around each point in the $(r_d,\Sigma_d)$-plane of data from the SPARC database and subsequently removing galaxies with negative values. On top of this -- to make sure that the deep MONDian regime is sampled well enough -- $500$ points are randomly generated at decreasing acceleration scales. Figure \ref{fig:1} illustrates both the original SPARC data (red) and the mock data (black) in the $(r_d,\Sigma_d)$-plane. 
For the analysis of mock galaxies, the Kuzmin disk is considered as a case study. For this mass distribution, MG has an exact analytic solution~\citep{Brada:1994pk}, so the exact difference between MI and MG can be extrapolated beyond the deep MONDian regime. The surface mass density of the Kuzmin disk is given by
\begin{equation}
\Sigma_d(r)=\Sigma_{0}\bigg(1+\frac{r^2}{r_d^2}\bigg)^{-\frac{3}{2}},
\end{equation} 
where $\Sigma_{0}$ denotes the central surface density and $r_d$ denotes the disk scale length. Additional mass distributions are considered in appendix \ref{app:mock}, using the Brada-Milgrom approximation for MG (Eq. $25$ in \citealt{Brada:1994pk}) which is valid for AQUAL-like theories. All these mass distributions give similar results as the Kuzmin disk.\newline
In order to asses when the DML is reached, an interpolation function must be specified for both MI and MG in the Bekenstein-Milgrom formulation. For example, for a step function, the DML is reached as soon as the maximum acceleration is less than $a_0$, but this is clearly an unrealistic situation. Thus, we use the (inverse) interpolation function corresponding to the radial acceleration relation (RAR,~\citealt{McGaugh:2016leg}); 
\begin{equation}
\nu_{RAR}(x)=\frac{1}{1-e^{-\sqrt{x}}}.
\label{int}
\end{equation}

\subsection{The deep MONDian limit (DML)}
\begin{figure*}[!h]
	\centering
	\captionsetup{width=1\textwidth}
	\includegraphics[width=0.42\textwidth]{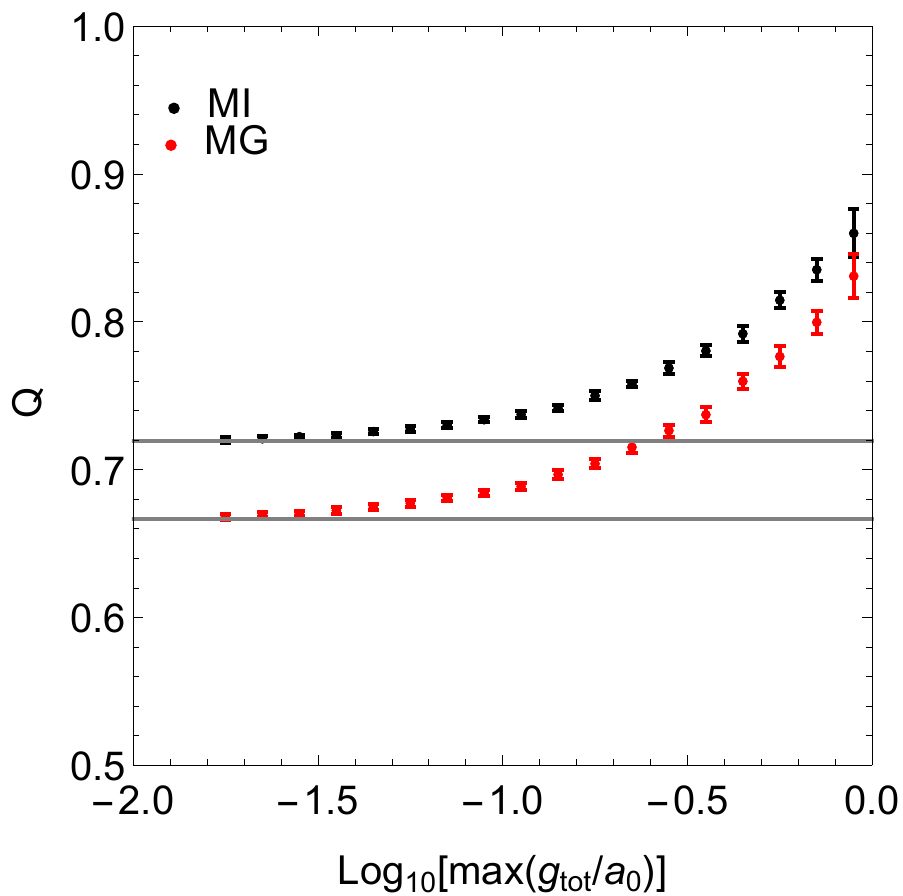}
	\includegraphics[width=0.45\textwidth]{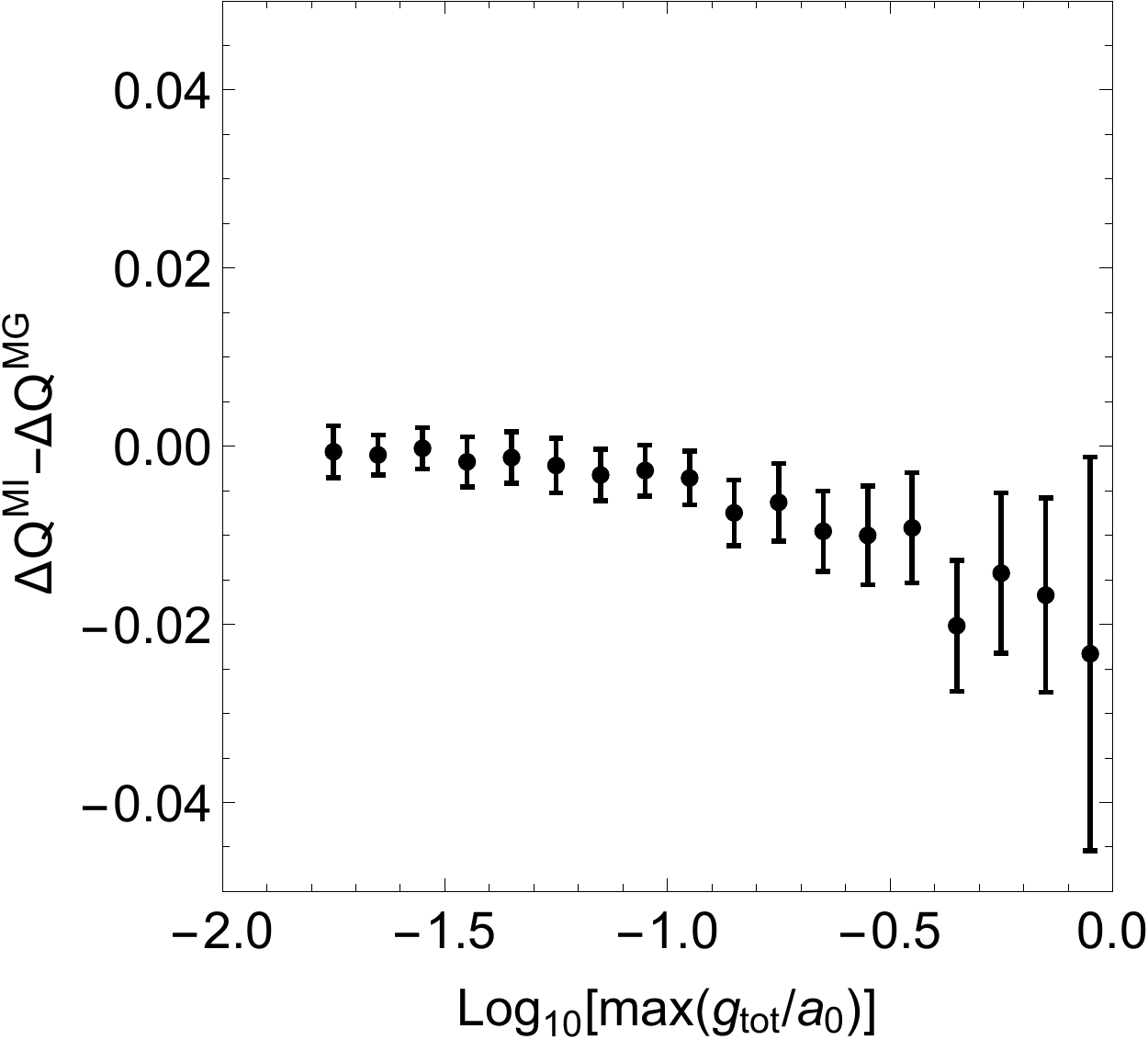}
	\caption{Left panel: $Q$ calculated for both MI and MG for a Kuzmin disk using mock data binned into acceleration bins. The gray lines denote the deep MONDian limits for MI and MG, respectively. Right panel: $\Delta Q^{MI}-\Delta Q^{MG}$ as a function of acceleration scale.}
	\label{fig:ti}
\end{figure*}

To gauge when the DML is reached, $Q$ is calculated from the mock data within bins of maximum total acceleration ($\max(g_{obs})$) for MI and MG. This is shown in the left panel of Figure \ref{fig:ti}, from which it is clear that the expected DML value of $Q$ is reached for $\max(g_{obs})\lesssim 0.01a_0$. These are extremely low accelerations: known disk galaxies are never entirely in such a deep MOND regime. However, the predictions for $Q$ for MI and MG drift in a similar way as $\max(g_{obs})$ increases. As long as the drifts of MI and MG are approximately equal, the measured $Q$ can -- in the context of differentiating between MI and MG -- be rescaled such that galaxies above the deep MONDian regime yield "correct" DML values of $Q$. Denote the drifts in MI and MG viz
\begin{equation}
\begin{split}
&\Delta Q^{MI}\equiv Q^{MI}-Q_{exact}^{MI},\\
&\Delta Q^{MG}\equiv Q^{MG}-Q_{exact}^{MG},\\
\end{split}
\end{equation} 
where $Q_{exact}^{MI}\approx 0.719$ and $Q_{exact}^{MG}=\frac{2}{3}$ denotes the exact values of $Q$. As long as $\Delta Q^{MI}-\Delta Q^{MG}\simeq 0$, $Q$ can be rescaled to the deep MONDian value with either $\Delta Q^{MI}$ or $\Delta Q^{MG}$. The right panel of Figure \ref{fig:ti} illustrates the difference $\Delta Q^{MI}-\Delta Q^{MG}$. $|\Delta Q^{MI}-\Delta Q^{MG}|\lesssim 0.01\simeq const$ for $\max_r(g_{obs})\lesssim 0.4 a_0$, meaning that the measured $Q^m$ can be corrected with $\Delta Q^{(DML)}=\Delta Q^{MI}\simeq \Delta Q^{MG}$ for galaxies which fulfill $\max(g_{obs})\lesssim 0.4 a_0$.
Table \ref{table:1} lists $\Delta Q^{(DML)}$ for different acceleration bins.
\begin{table}[H]
	\caption{Acceleration scale corrections. The uncertainty on $\Delta Q^{(DML)}$ is negligible.} % title of Table
	\label{table:1} % is used to refer this table in the text
	\centering % used for centering table
	\begin{tabular}{c c} % centered columns (4 columns)
		\hline\hline % inserts double horizontal lines
		$\log_{10}[\max(g_{obs}/a_0)]$ & $\Delta Q^{(DML)}$  \\
		\hline % inserts single horizontal line
		$[-1.8,-1.4]$&$0.00$ \\
		$[-1.4,-1.1]$&$0.01$\\
		$[-1.1,-0.8]$&$0.02$\\
		$[-0.8,-0.7]$&$0.03$\\
		$[-0.7,-0.6]$&$0.04$\\
		$[-0.6,-0.5]$&$0.05$\\
		$[-0.5,-0.4]$&$0.06$\\
		\hline %inserts single line
	\end{tabular}
\end{table}

\subsection{Range of sampling}
\begin{figure*}[!ht]
	\centering
	\captionsetup{width=1\textwidth}
	\includegraphics[width=0.42\textwidth]{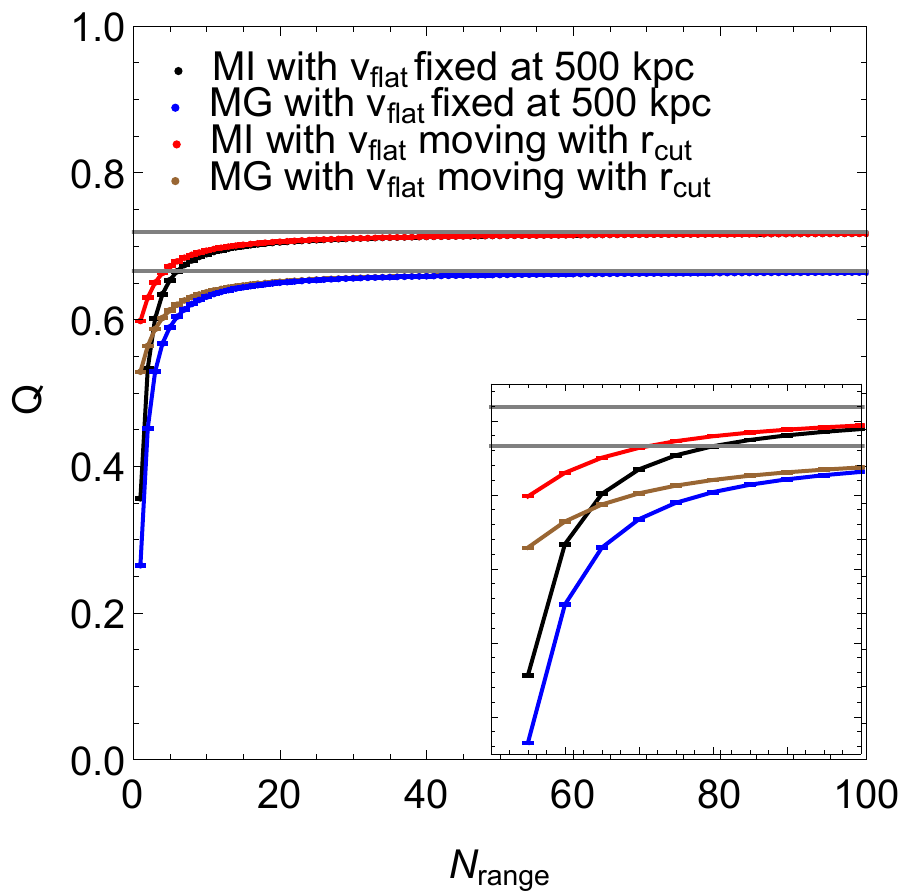}
	\includegraphics[width=0.45\textwidth]{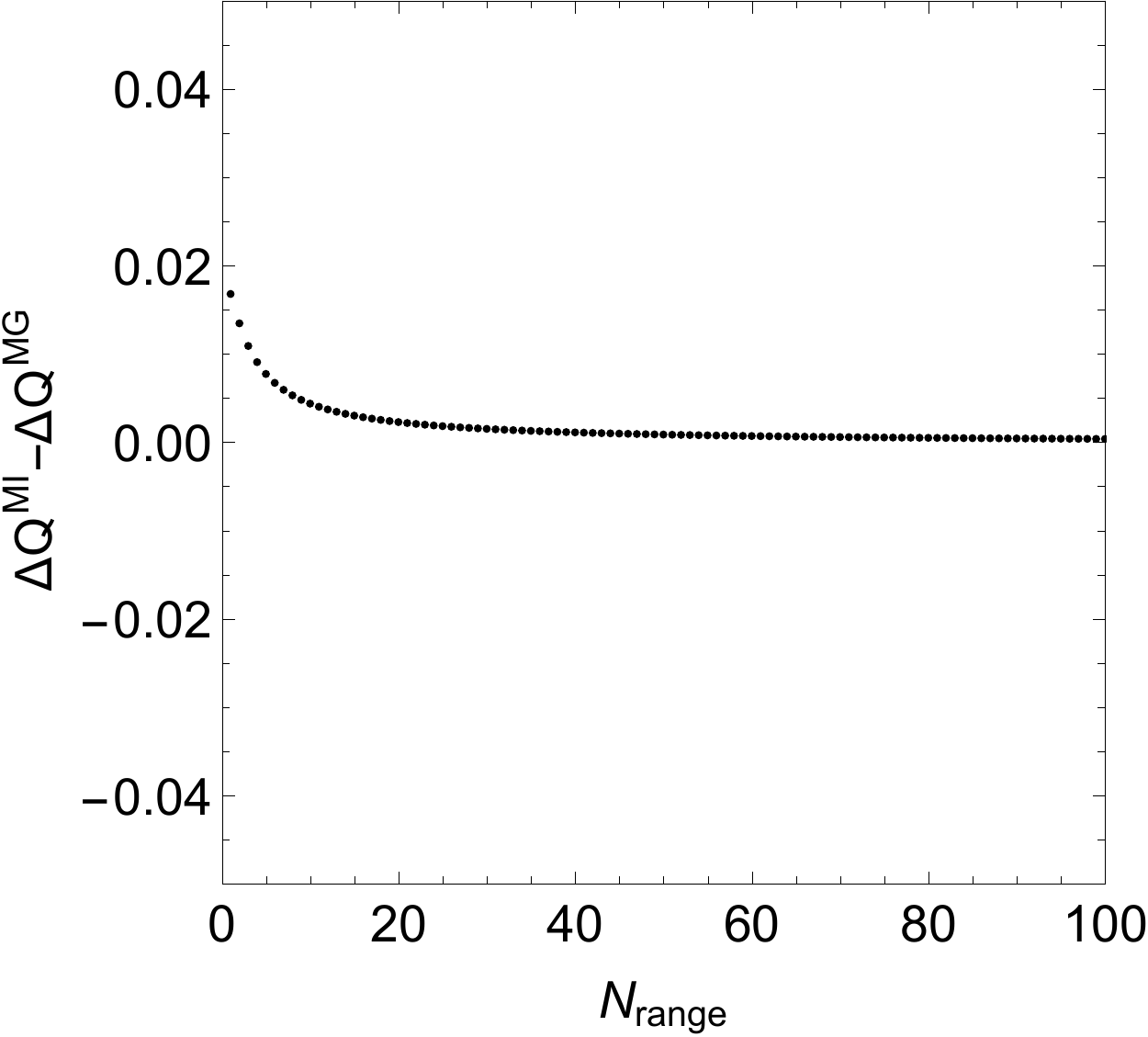}
	\caption{Left panel: Arithmetic mean of $Q$ for all mock galaxies with $\max[g_{obs}]\leq 0.01 a_0$ as a function of radial range in units of $r_d$. Black and blue lines represent MI and MG, respectively, with $v_{\infty}\simeq v_{flat}=v_{tot}(500kpc)$. Cyan and brown lines represent MI and MG, respectively, with $v_{\infty}\simeq v_{flat}=v_{tot}(r_{cut})$, where $r_{cut}$ denotes the radii at which the integrals are cut. The inset is a magnification of $N_{range}=[0,10]$. Right panel: $\Delta Q^{MI}-\Delta Q^{MG}$ as a function of radial range. The error bars are smaller than the points.}
	\label{fig:3f}
\end{figure*}

To gauge the effect of the finite extension of galaxy rotation curves and mass surface density profiles, $Q$ is computed at decreasing multiples of the disk scale length ($r_{cut}=N_{range} r_d$). The left panel of Figure \ref{fig:3f} illustrates the arithmetic mean of $Q$ of the mock data for which $\max[g_{obs}]\leq 0.01 a_0$ as a function of decreasing $N_{range}$. $Q$ is shown for both MI and MG calculated in two different ways: i) with $v_{\infty}$ fixed at $r=500 kpc$ to isolate the effect of neglecting part of the integral in the numerator of $Q$ and ii) $v_{\infty}$ approximated by $v_{tot}(r_{cut})$ such that it provides a better representation of the observational situation. 
From the left panel of Figure \ref{fig:3f} it is clear that all curves approach the theoretical values at increasing $r_{cut}$. At small ranges $Q$ decreases for both approximations of $v_\infty$. However, with the moving approximation of $v_\infty$, $Q$ decreases significantly less than the fixed one (for both MI and MG). This is a lucky occurrence indicating that the underestimate of $Q$ from cutting down the integral is somewhat counter-balanced by the underestimate of $v_\infty$ from the finite size of galaxy disks.  Indeed, for galaxies with $\max[g_{obs}] < 0.01 a_0$, the rotation curve is still slowly rising up to $500 kpc$, so $v_{obs}(r_{cut}) < v_\infty$. Hence, the measured value of $Q$ is closer to the theoretical value in the actual observational situation, since we can only estimate $v_{flat}$ near $r_{cut}$.\newline

\noindent From the left panel of Figure \ref{fig:3f} it is clear that MI and MG drift similarly as the range is decreased. As long as the drifts of MI and MG are approximately equal, the measured $Q$ can be corrected similarly to the case of the acceleration scale in the previous section. The right panel of Figure \ref{fig:3f} shows $\Delta Q^{MI}-\Delta Q^{MG}$ in the context of varying range. $|\Delta Q^{MI}-\Delta Q^{MG}|\lesssim 0.01\simeq const$ for $N_{range}\gtrsim 3$, meaning that the measured $Q$ can be corrected with $\Delta Q^{(range)}=\Delta Q^{MI}\simeq \Delta Q^{MG}$ for galaxies which fulfill $\max[r]\gtrsim 3r_d$. Table \ref{table:2} lists $\Delta Q^{(range)}$ for different ranges. The approach to the theoretical limit for increasing range is slow for the Kuzmin disk. For this reason $\Delta Q^{(range)}=0$ is only reached for $N_{range}>55$ in this case. The corrections in Table \ref{table:2} isolate the range effect from the acceleration effect. To mimick the observational situation, however, we have repeated the same exercise considering galaxies with $\max(g_{obs})<0.15a_0$, applying the corrections in Table \ref{table:1}. We find that the corresponding corrections due to $N_{range}$ are very similar to those given in Table \ref{table:2}.
\begin{table}[H]
	\caption{Range corrections. The uncertainty on $\Delta Q^{(range)}$ is negligible.} % title of Table
	\label{table:2} % is used to refer this table in the text
	\centering % used for centering table
	\begin{tabular}{c c} % centered columns (4 columns)
		\hline\hline % inserts double horizontal lines
		$N_{range}$ & $\Delta Q^{(range)}$\\
		\hline % inserts single horizontal line
		$[18,55]$ & $ 0.01 $ \\
		$[11,18]$ & $ 0.02 $\\
		$[8,11]$ & $0.03$ \\
		$[6,8]$ & $0.04$\\
		$5$ & $0.05$ \\
		$4$ & $0.06$ \\
		$3$ & $0.07$\\
		\hline %inserts single line
	\end{tabular}
\end{table}

\subsection{Sampling resolution}
\begin{figure*}[!ht]
	\centering
	\captionsetup{width=1\textwidth}
	\includegraphics[width=0.47\textwidth]{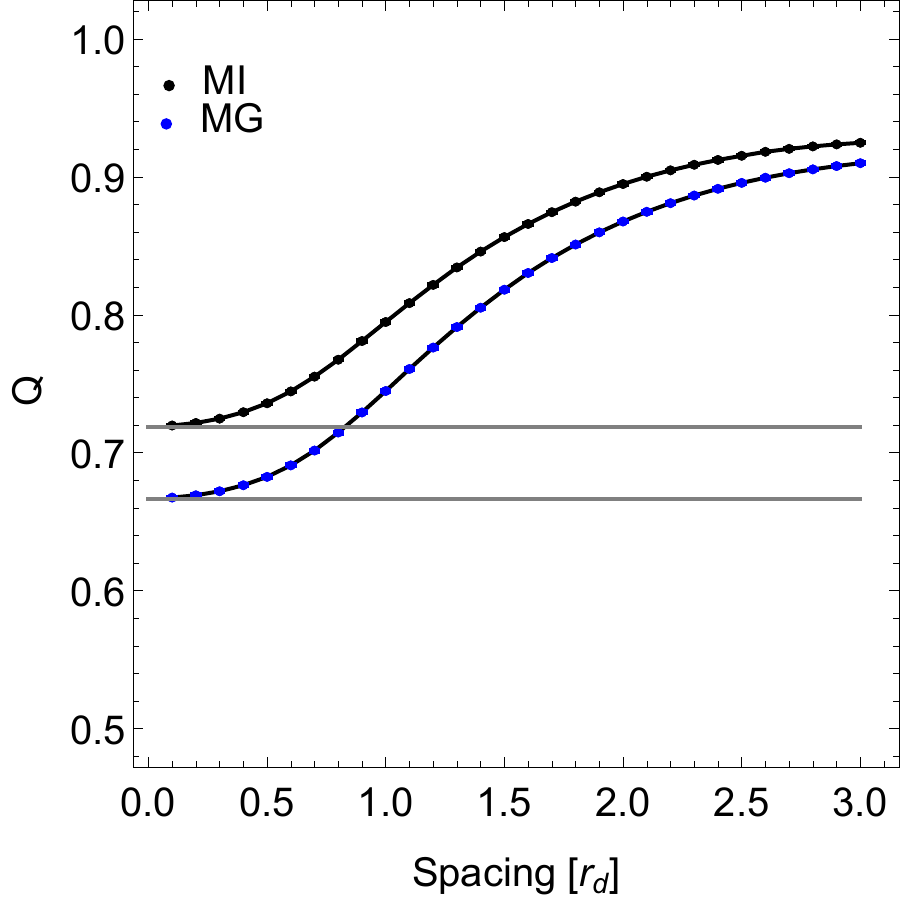}
	\includegraphics[width=0.5\textwidth]{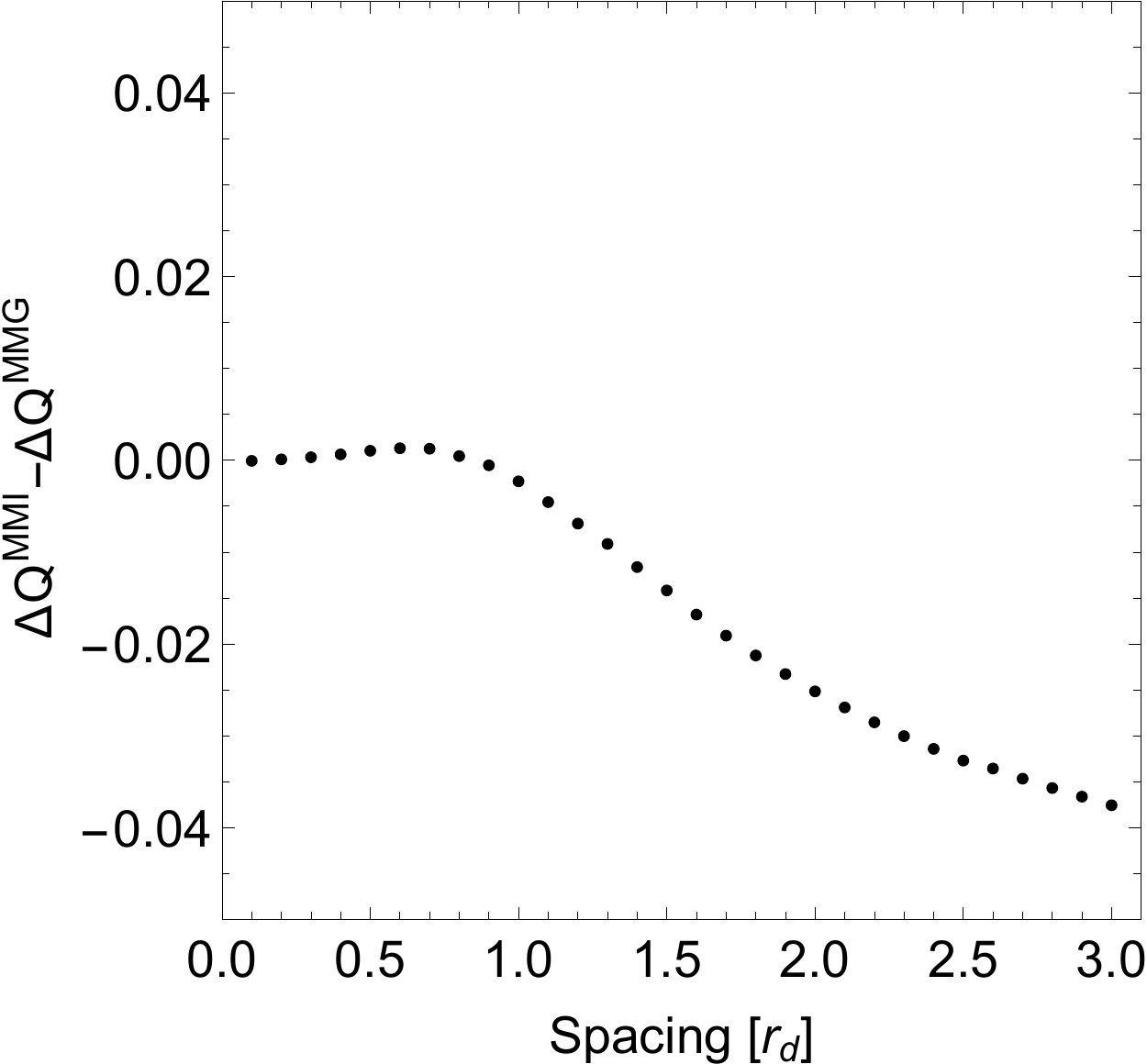}
	\caption{Left panel: Arithmetic mean of $Q$ for all mock galaxies with $\max[g_{obs}]\leq 0.01 a_0$ as a function of spacing between sampled points. Black and blue lines represent MI and MG, respectively. Right panel: $\Delta Q^{MI}-\Delta Q^{MG}$ as a function of spacing between sampled points.}
	\label{fig:4f}
\end{figure*}

In order to gauge the effect of the spatial resolution of the observations, $Q$ is calculated as a function of the spacing between sampled points in units of $r_d$. The sampling for each galaxy is performed in the range $r_{j}\in [0.01,200]r_d$. In order to compute $Q$ from a discrete set of points, the integral of equation \eqref{q1} is discretized viz
\begin{equation}
Q\simeq\frac{S_1+S_2}{v_{f}^2\big[S_3+S_4\big]},
\label{q3}
\end{equation}
with
\begin{equation}
\begin{split}
&S_1\equiv \sum_{j=1}^{N_-1}(r_{j}-r_{j+1})r_{j}\Sigma_d(r_j)v_{tot}^2(r_j),\\
&S_2\equiv \sum_{j=2}^{N}(r_{j-1}-r_{j})r_{j}\Sigma_d(r_j)v_{tot}^2(r_j),\\
&S_3\equiv \sum_{j=1}^{N_-1}(r_{j}-r_{j+1})r_{j}\Sigma_d(r_j), \qquad \text{and}\\
&S_4\equiv \sum_{j=2}^{N}(r_{j-1}-r_{j})r_{j}\Sigma_d(r_j),\\
\end{split}
\end{equation}
where $v_f\simeq v_\infty$ being the asymptotically flat velocity which is determined as the arithmetic mean between the chain of points that are within $5\%$ of the arithmetic mean of the two outermost points (similarly to \citealt{Lelli:2016zqa}). If the third outermost point is not within $5\%$ of the arithmetic mean of the two outermost points, the galaxy is not assigned an asymptotically flat velocity.\newline 
The left panel of Figure \ref{fig:4f} shows the arithmetic mean of $Q$ calculated via Equation \eqref{q3} -- as predicted by MI (black) and MG (blue) -- of the mock data for which $\max(g_{obs})\leq 0.01 a_0$ as a function of the spacing between the sampled points in units of $r_d$. As expected, $Q$ approaches the theoretical value for extremely small spacings ($r < 0.1 r_d$) and progressively increases for larger spacing between the sampled points.\newline
From the right panel of Figure \ref{fig:4f} it is clear that the increase for MI and MG are similar, meaning that -- similarly to the acceleration scale and range cases -- the measured $Q$ can be rescaled such that galaxies with spacings larger than $\sim 0.5r_d$ yield $Q$ approximately equal to that calculated with infinitely small spacing.
$|\Delta Q^{MI}-\Delta Q^{MG}|\lesssim 0.01\simeq const$ for spacing $\gtrsim 1.3r_d$, meaning that the measured $Q$ can be corrected with $\Delta Q^{(res.)}=\Delta Q^{MI}\simeq \Delta Q^{MG}$ for galaxies with spacing $\lesssim 1.3r_d$. Table \ref{table:3} lists $\Delta Q^{(res.)}$ for different spacing between sampled points.
\begin{table}[H]
	\caption{Spacing corrections. The uncertainty on $\Delta Q^{(res.)}$ is negligible.} % title of Table
	\label{table:3} % is used to refer this table in the text
	\centering % used for centering table
	\begin{tabular}{c c} % centered columns (4 columns)
		\hline\hline % inserts double horizontal lines
		Spacing $[r_d]$ & $\Delta Q^{(res.)}$  \\
		\hline % inserts single horizontal line
		$1.3$ &$ 0.12 $\\
		$1.2$ &$ 0.10 $\\
		$1.1$ &$ 0.09$\\
		$1.0$ &$ 0.08 $\\
		$0.9$ &$ 0.06 $\\
		$0.8$ &$ 0.05 $\\
		$0.7$ &$ 0.04 $\\
		$0.6$ &$ 0.03 $\\
		$0.5$ &$ 0.02 $\\
		$0.4$ &$ 0.01 $\\
		$0.3$ &$ 0.01$\\
		$0.2$ &$ 0.00 $\\
		\hline %inserts single line
	\end{tabular}
\end{table}

\section{Application to SPARC galaxies}
\label{sec:Qmes}
We will now use actual galaxies from the SPARC database~\citep{Lelli:2016zqa} to calculate $Q$. The SPARC database consists of $175$ rotationally supported galaxies spanning broad dynamic ranges in stellar mass, surface brightness and gas fraction. The database provide observed rotational velocities ($v_{obs}$), along with the associated uncertainties ($\delta v_{obs}$), as well as distance ($D$) and inclination ($i$) measurements for each galaxy. The database also provides the rotational velocities due to the baryonic components, which are computed using the Spitzer $[3.6]$ surface brightness profile for the stars and the HI surface density profile for the gas. Finally, the database provides the central surface brightness ($\mu_d$) and exponential scale length ($r_d$) of the stellar disk (see Figure \ref{fig:1}). In this paper, however, we estimated the scale length of the baryonic disk (gas plus stars) by fitting an exponential profile to the total surface mass density profile of both stars and gas (see appendix \ref{app:rot}).\newline\newline
Following \citet{Lelli:2017vgz}, $12$ galaxies are discarded from the analysis based on a quality flag (see \citealt{Lelli:2016zqa}). We note that in this work we keep face-on galaxies ($i<30^\circ$) because $Q$ does not depend on $i$.\newline

\noindent The baryonic velocity is computed viz (recall we are only considering disk-only galaxies) 
\begin{equation}
	v_{bar}(r)=\sqrt{|v_{g}(r)|v_{g}(r)+\Upsilon^{d}v_{d}^2(r)},
	\label{v_bar}
\end{equation}
where $v_g$ is the gas contribution, $v_d$ is the stellar contribution, and $\Upsilon_d$ is the stellar mass-to-light ratio. In line with \citet{Mcagugh:2014,Lelli:2016zqa} $\Upsilon^{d}=0.5\frac{M_{\odot}}{L_\odot}$ is taken with a $25 \%$ uncertainty. \newline\newline

\begin{figure*}[!ht]
	\centering
	\captionsetup{width=1\textwidth}
	\includegraphics[width=0.35\textwidth]{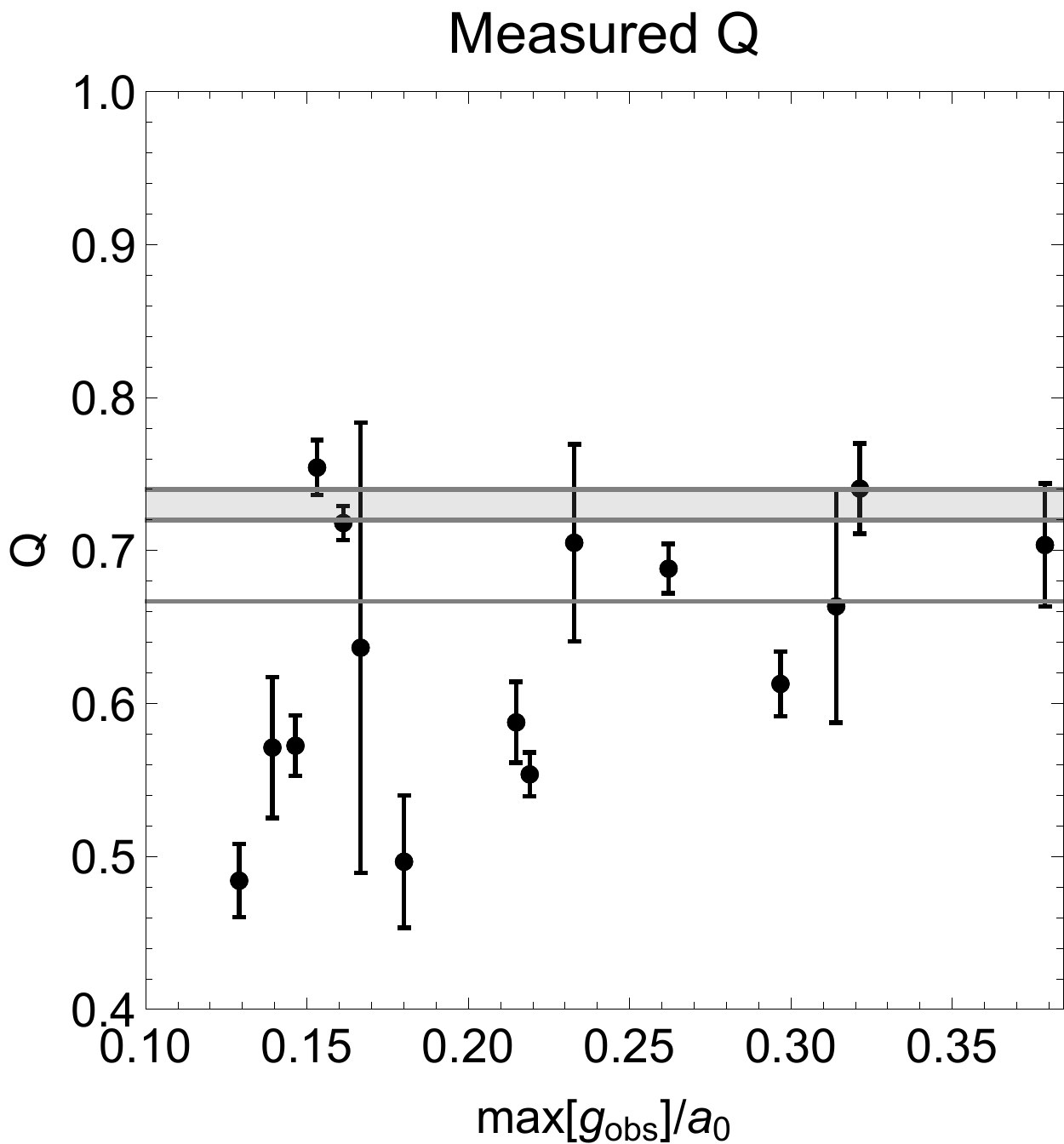}
	\includegraphics[width=0.35\textwidth]{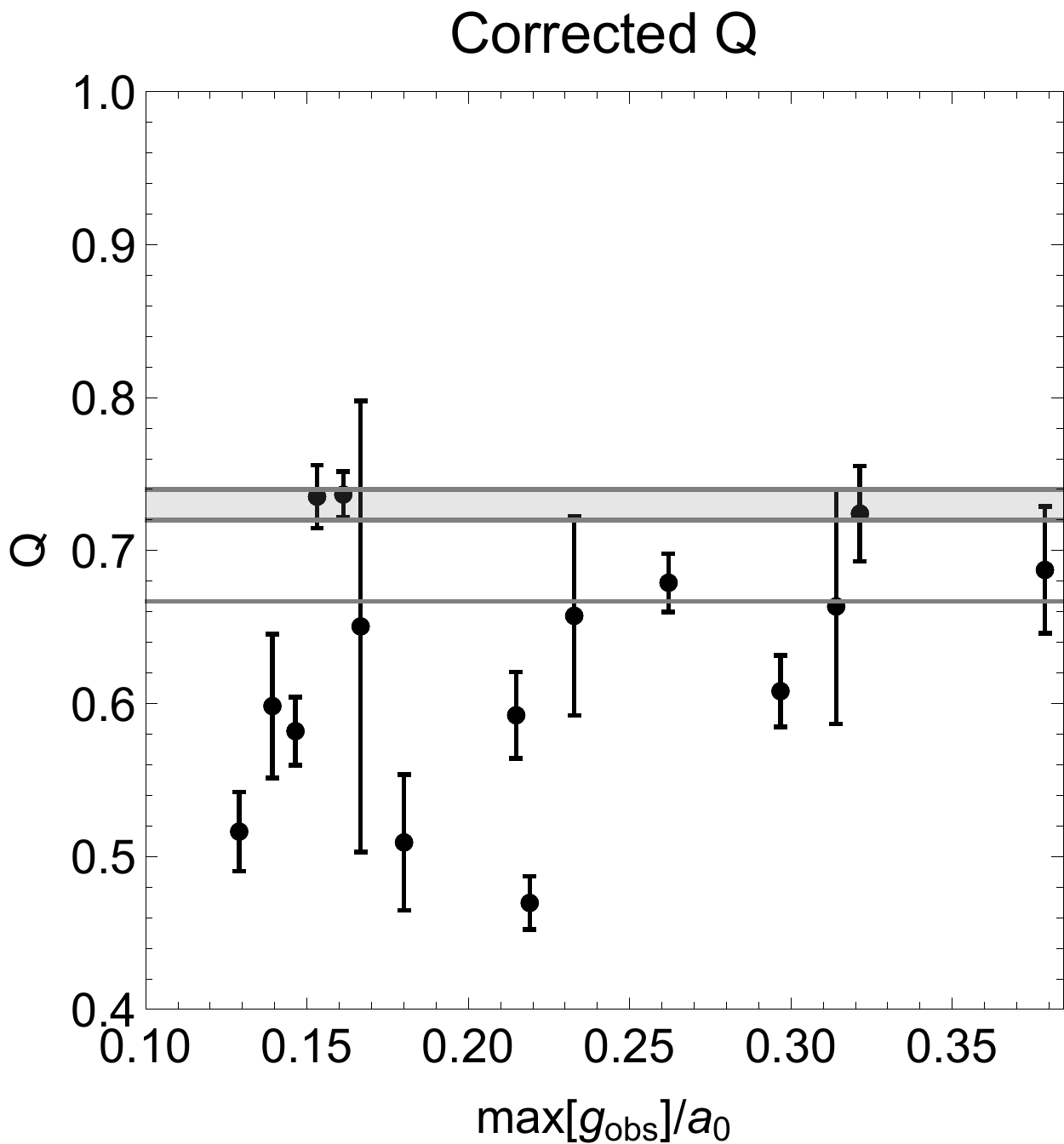}\\
	\includegraphics[width=0.35\textwidth]{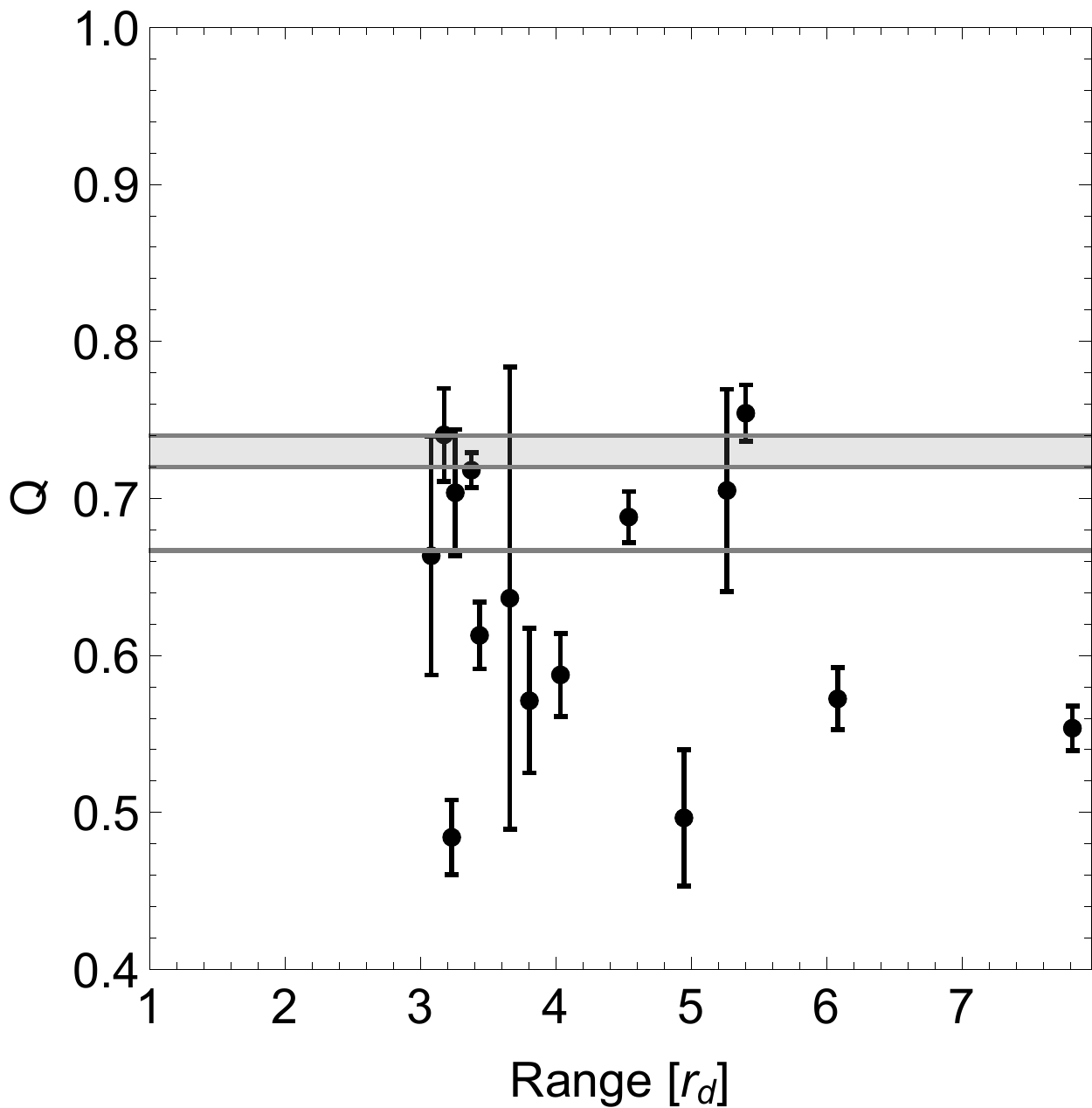}
	\includegraphics[width=0.35\textwidth]{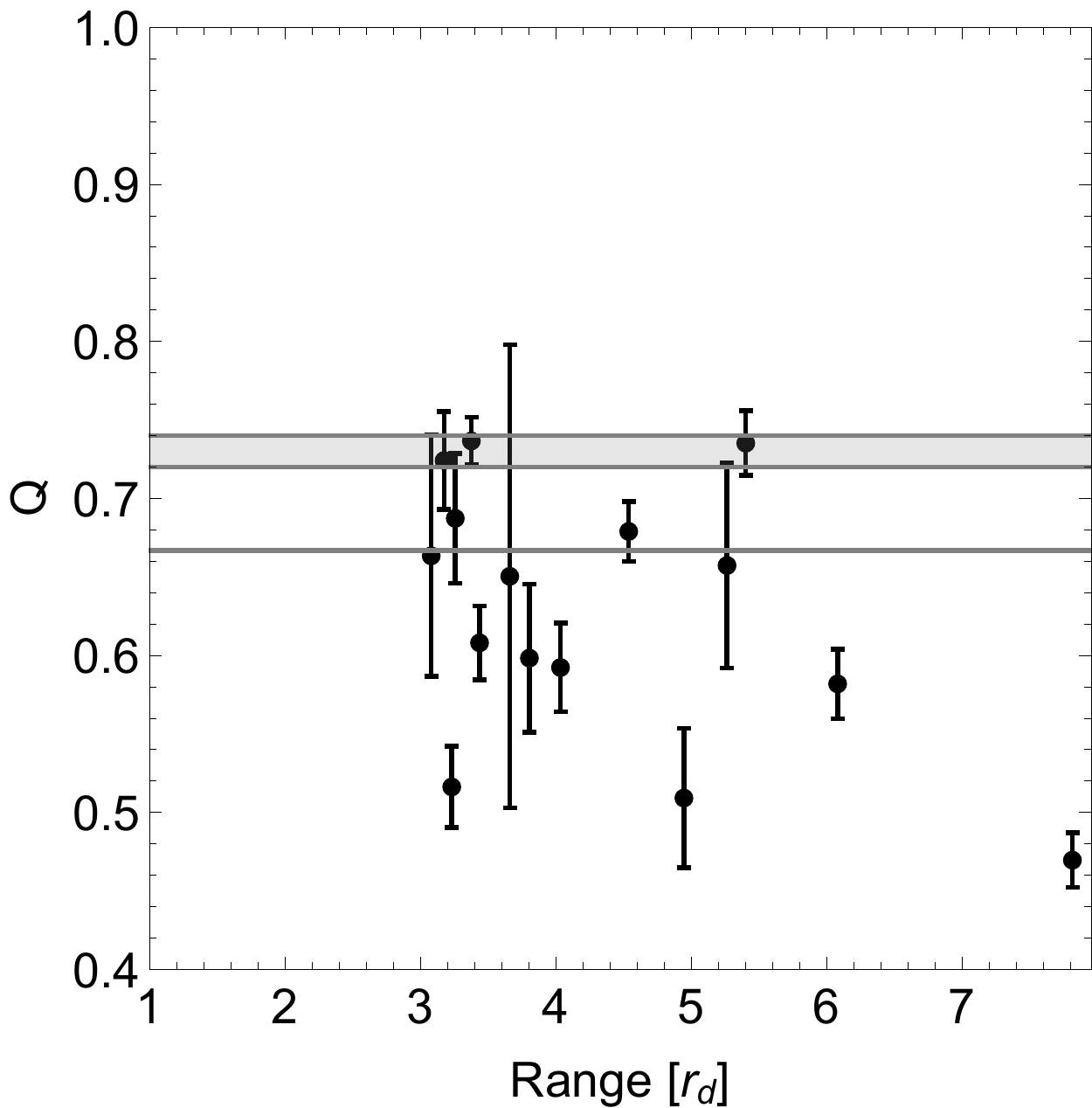}\\
	\includegraphics[width=0.35\textwidth]{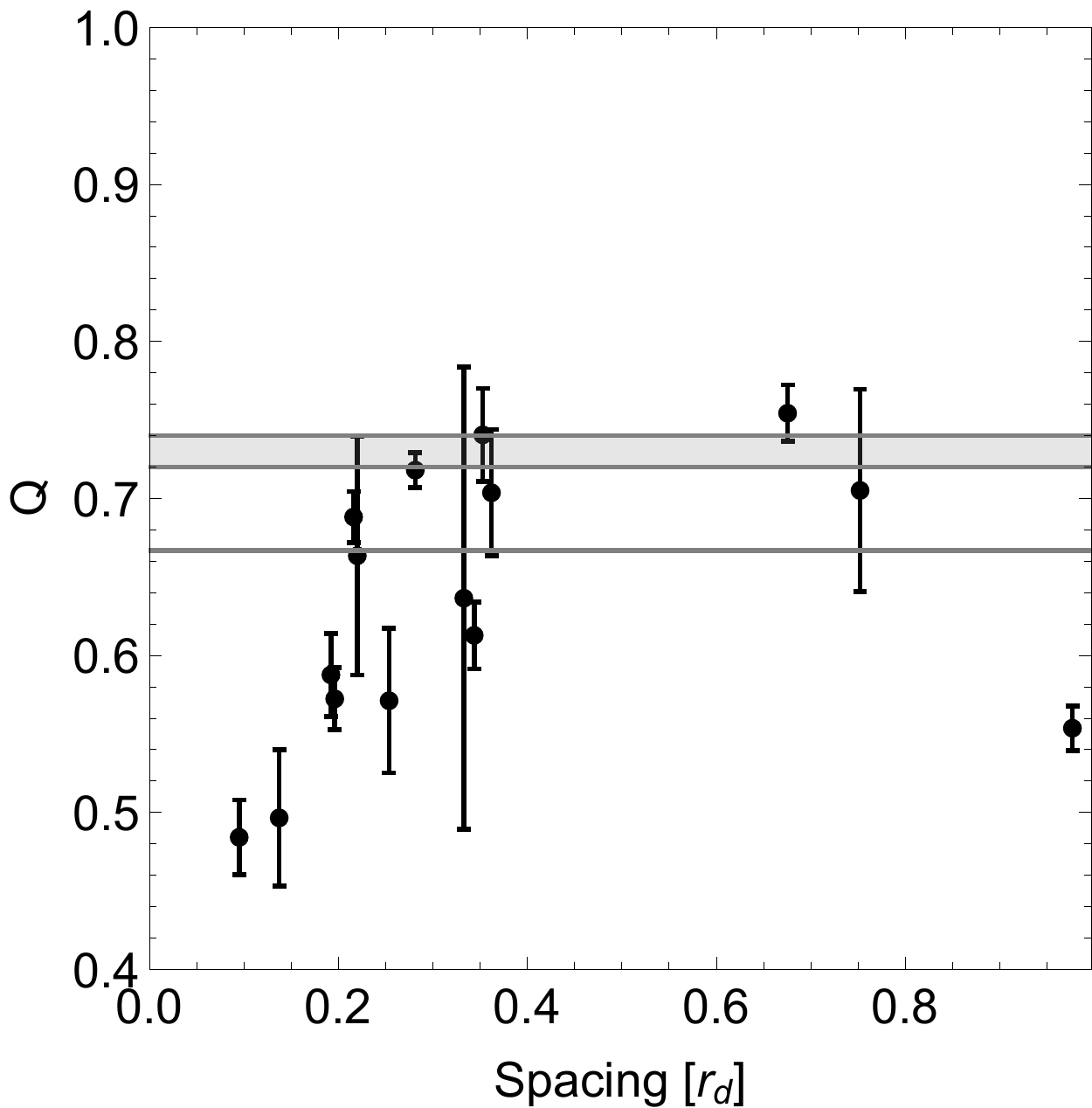}
	\includegraphics[width=0.35\textwidth]{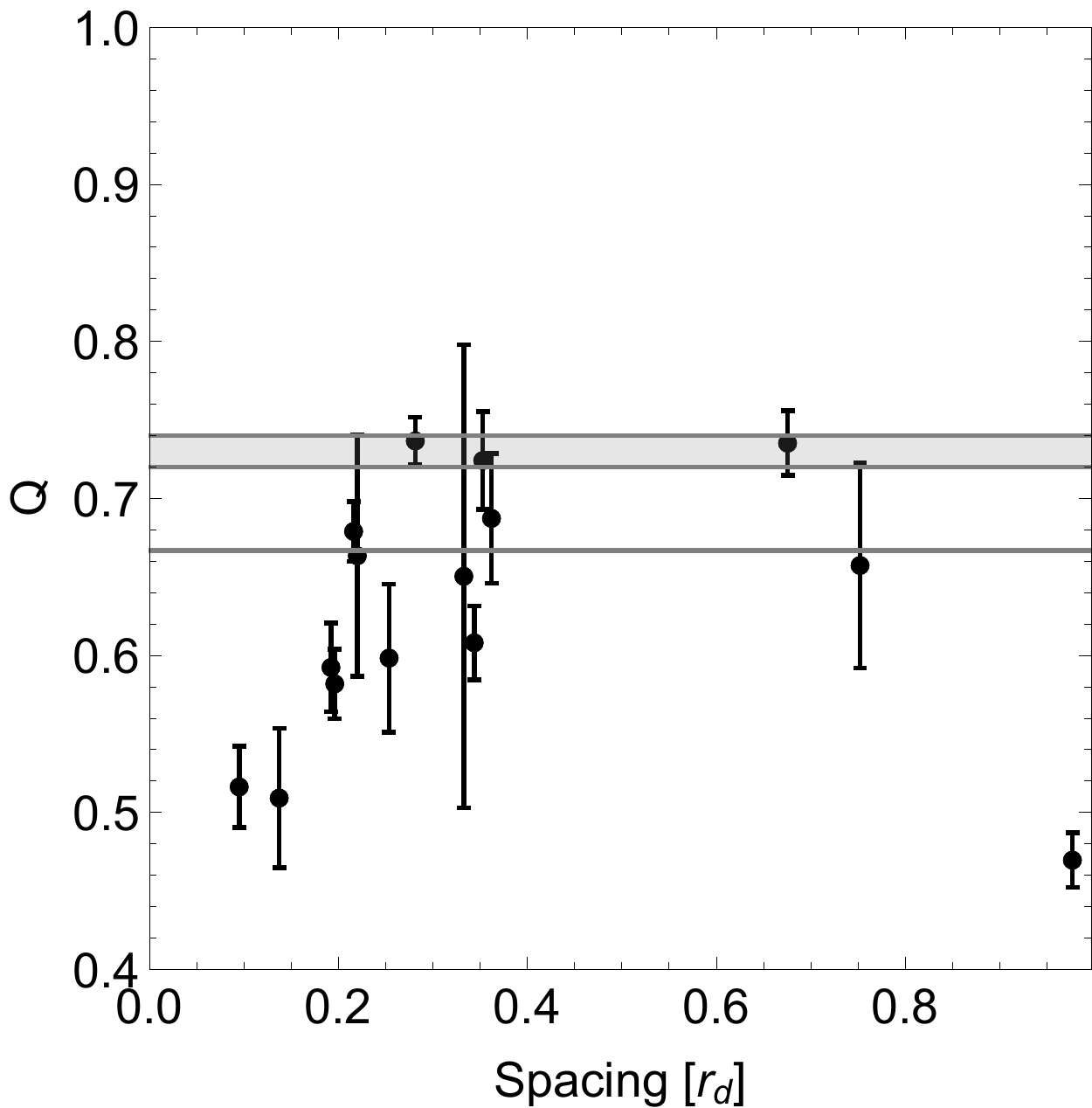}
	\caption{$Q$ for individual galaxies as a function of $\max[g_{obs}]$ (first row), $\frac{\max[r]}{r^d}$ (second row) and spacing (third row). These $15$ galaxies are deep enough in the MOND regime and have sufficiently extended and finely sampled rotation curves (see section \ref{app:mock}). The gray line shows the predicted $Q$ from MOND modified gravity, while the gray band shows the ranges of values expected in MOND modified inertia.}
	\label{fig:re}
\end{figure*}

\noindent $Q$ is discretized as shown in Equation \eqref{q3}, however with $\Sigma_d\rightarrow \mu_{d}\Upsilon^d+ 1.33\Sigma_{HI}$. Requiring that there be a measured gas profile (approximated by the $HI$ profile) and a well defined value for $v_f$ on top of the accelerations scale, range and resolution requirements yields $15$ galaxies from the SPARC database. In appendix \ref{app:rot}, we present the rotation curves for these galaxies as well as a table detailing $Q^{(m)}, Q^{(c)}$, the acceleration scale, the range and the spacing for each galaxy. Figure \ref{fig:re} shows $Q$ as a function of the acceleration scale ($\max[g_{obs}]$), the range and the resolution. The left column of the Figure shows $Q^{(m)}$ whereas the right column shows $Q^{(c)}$.
The averages are calculated viz
\begin{equation}
\braket{Q_i}=\frac{1}{N}\sum_{i\in gal}Q_i\pm \sqrt{\frac{1}{N-1}\sum_{i\in gal}\bigg(\frac{1}{N}\sum_{j\in gal}Q_j-Q_i\bigg)^2},
\end{equation} 
where $N=\sum_{j\in gal}1$, with $gal$ denoting the relevant galaxies, is the number of measurements from a galaxy. The median of $Q$ from individual galaxies can be written
\begin{equation}
\braket{Q_i}=\text{Median}[Q_i]\pm \text{Median}\big[\big|Q_i-\text{Median}[Q_i]\big|\big].
\end{equation}
From all three rows there is a slight tendency of the average $Q$ decreasing as a the requirements are fulfilled to a higher degree. Let $\braket{Q^{(m)}}$ denote a list of the arithmetic mean and the median of $Q^m$, respectively. From the galaxies shown in Figure \ref{fig:re} $\braket{Q^m}$ is given by
\begin{equation}
\braket{Q^{(m)}}=\{0.63\pm 0.09, 0.64 \pm 0.07 \}.
\label{qe}
\end{equation}
For the corrected $Q$
\begin{equation}
\braket{Q^{(c)}}\approx\{0.63\pm 0.08, 0.65 \pm 0.06 \}.
\label{qe1}
\end{equation}
Comparing Equation \eqref{qe} and \eqref{qe1} it is clear that i) the arithmetic mean and median yield similar results and ii) the \emph{sum} of corrections have a very little impact on the two measures of the average $Q$, meaning that the corrections approximately cancel out. \newline

Both the mean and median values of $Q^m$ and $Q^c$ are consistent with the MOND predictions within $1.5$ sigma. With the current level of accuracy, we cannot distinguish between MG and MI, albeit there is a slight preference for the lower MG value. We also note that individual galaxies can significantly deviate from the predicted $Q$ values (at more than $5\sigma$), but this is likely due to the fact that the errorbars on a single object cannot possibility account for all the systematic uncertainties. A larger galaxy sample would have allowed us to determine whether $Q$ converges toward a characteristic value and to rigorously estimate the corresponding error. We expect that a sample of $\sim100-200$ galaxies satisfying our quality criteria should be sufficient to carry out this experiment. This may be possible in the near future thanks to large HI surveys with the Square Kilometer Array (SKA) and its pathfinders (e.g., \citealt{duffy2012}). \newline

\noindent The effect of the stellar mass-to-light ratio on the measured $Q$ is relatively minor in these $15$ galaxies. If we assume $\Upsilon^{d}=0.25\frac{M_{\odot}}{L_\odot}$ rather than $\Upsilon^{d}=0.5\frac{M_{\odot}}{L_\odot}$  we obtain a slightly different set of galaxies with different corrections since the value of $r_d$ of each galaxy slightly changes. Then Eqs. \eqref{qe} and \eqref{qe1} become
\begin{equation}
\braket{Q^{(m)}}=\{0.65\pm 0.09, 0.68 \pm 0.08 \},
\end{equation}
\begin{equation}
\braket{Q^{(c)}}\approx\{0.65\pm 0.09, 0.67 \pm 0.07 \}.
\end{equation}
These values are consistent with the previous ones within the errors, but they are systematically higher by $\sim0.02-0.04$ and become closer to the predicted value from MI.\newline
In \citet{Li:2018tdo} the RAR is fitted to individual galaxies yielding $\Upsilon^d$ values that are maximally favorable for MI. The $Q$ values corresponding to Eqs. \eqref{qe} and \eqref{qe1} are in this case
\begin{equation}
\braket{Q^{(m)}}=\{0.65\pm 0.08, 0.67 \pm 0.07 \},
\end{equation}
\begin{equation}
\braket{Q^{(c)}}\approx\{0.65\pm 0.09, 0.66 \pm 0.06 \}.
\end{equation}
The $Q$ values are very close to the case of $\Upsilon^{d}=0.25\frac{M_{\odot}}{L_\odot}$. Intriguingly, even when we adopt M/L values that are maximally favorable for MI, the resulting $Q$ values remain closer to the predictions of MG.

\noindent Lastly we note that the analysis of $Q$ conducted in this paper is closely related to the fractions of accelerations investigated in \citet{Frandsen:2018ftj,Petersen:2019obe} since both quantities involve ratios between velocities and benefit from the cancellation of systematic uncertainties (inclination angle, distance and partially the mass to light ratio). However, despite the apparent similarities, the finer details in calculating $Q$ and the fractions of accelerations in \citet{Frandsen:2018ftj,Petersen:2019obe} lead to a different analysis and in the end the results cannot be compared.

\section{Summary and conclusions}
\label{sec:discussion}
In this paper, we present a first attempt to differentiate between MOND modified inertia (MI) and MOND modified gravity (MG) using galactic rotation curve data. In particular, we investigate $Q=\frac{\braket{v_{tot}^2}}{v_\infty^2}$ defined in \citet{Milgrom:2012rk}, for which MI predicts $Q=0.73\pm 0.01$ and MG predicts $Q=\frac{2}{3}$ for disk-only galaxies everywhere in the deep MONDian regime. We compare theoretical predictions to data from the SPARC database. To do so, we thoroughly discuss the impact of systematic uncertainties in the value of $Q$ via investigating a set of mock galaxies. Specifically we consider the impact on $Q$ of the acceleration scale ($\max[g_{obs}]$), the spacing between sampled points (resolution) and the range of sampling ($\max(r)$). We find that the systematic uncertainties in $Q$ can be approximately accounted for as long as $\max[g_{obs}]\lesssim 0.4a_0$ (acceleration scale requirement), $\max[r]\gtrsim3r^d$ (sampled range requirement) and the spacing between points is $\lesssim 1.3r_d$ (resolution requirement). Imposing these criteria on the SPARC database leaves $15$ galaxies. Before correcting for systematic effects, the arithmetic mean and median of $Q$ is given by $\braket{Q^{(m)}}=\{0.63\pm 0.09, 0.64 \pm 0.07 \}$, respectively. After correction $\braket{Q^{(c)}}\approx\{0.63\pm 0.08, 0.65 \pm 0.06 \}$. From $\braket{Q^{(m)}}$ and $\braket{Q^{(c)}}$ several things can be noted; i) the arithmetic mean and median yield similar predictions, ii) these measurements line up closely with the predictions of MG both before and after correction, although the prediction of MI is still within $1.5\sigma$.\newline 
Future HI surveys with the SKA and its pathfinders are expected to provide HI rotation curves for thousands of galaxies, which will allow us to achive more solid estimates of Q and push down the statistical errors.

\section*{Acknowledgements}
We thank the referee for a constructive report that improved the clarity of our paper. JP also thanks ESO for the hospitality that made this collaboration possible as well as the partial funding from The Council For Independent Research, grant number DFF 6108-00623. The CP3-Origins center is partially funded by the Danish National Research Foundation, grant number DNRF90.

\appendix
	\section{Mock galaxies for different mass distributions}
	\label{app:fig}
\begin{figure*}[ht]
		\centering
		\captionsetup{width=1\textwidth}
		\includegraphics[width=0.24\textwidth]{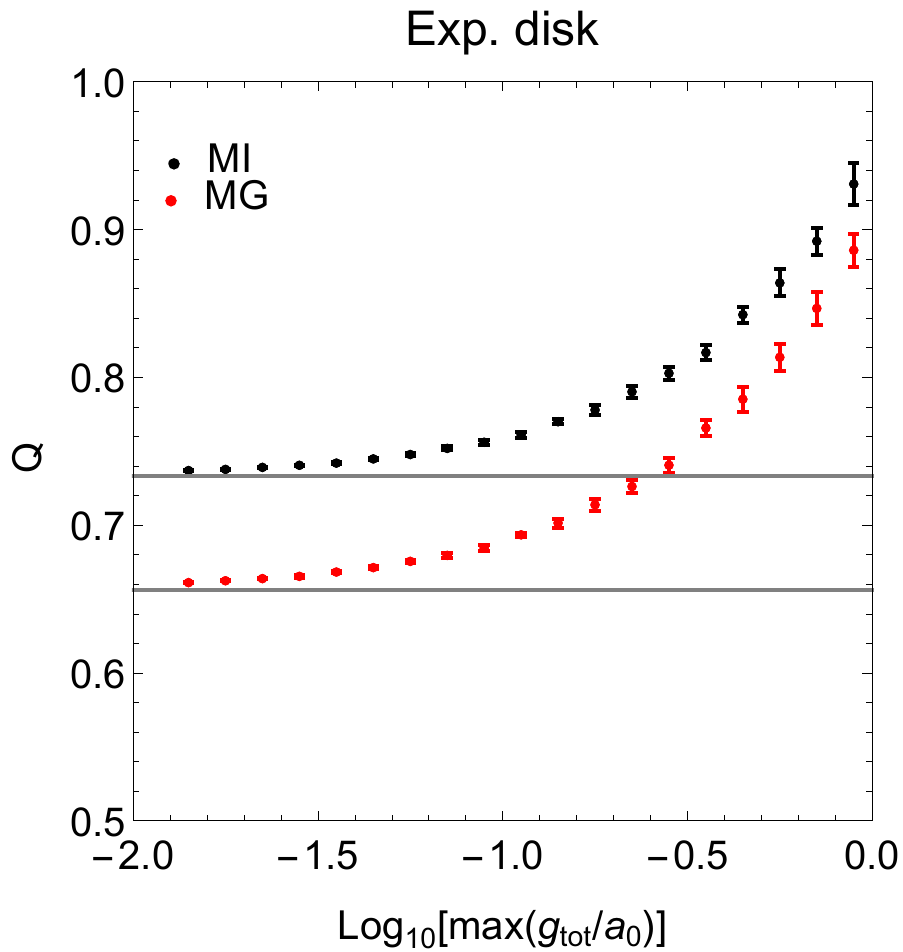}
		\includegraphics[width=0.24\textwidth]{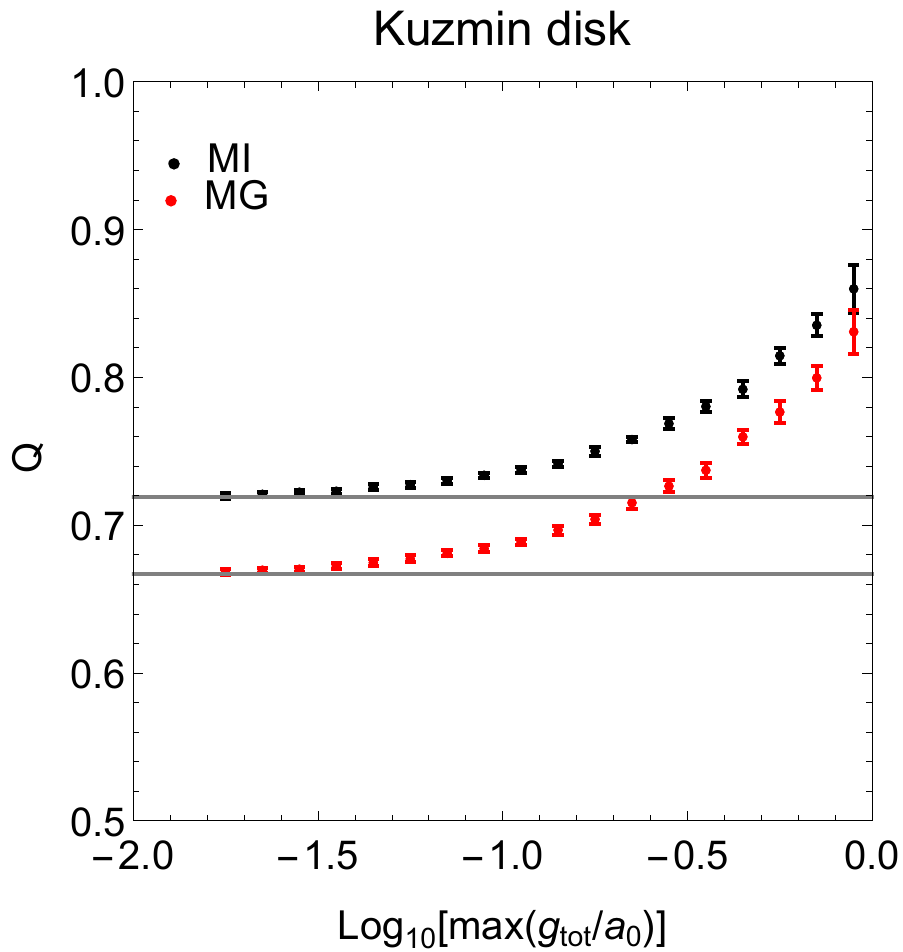}
		\includegraphics[width=0.24\textwidth]{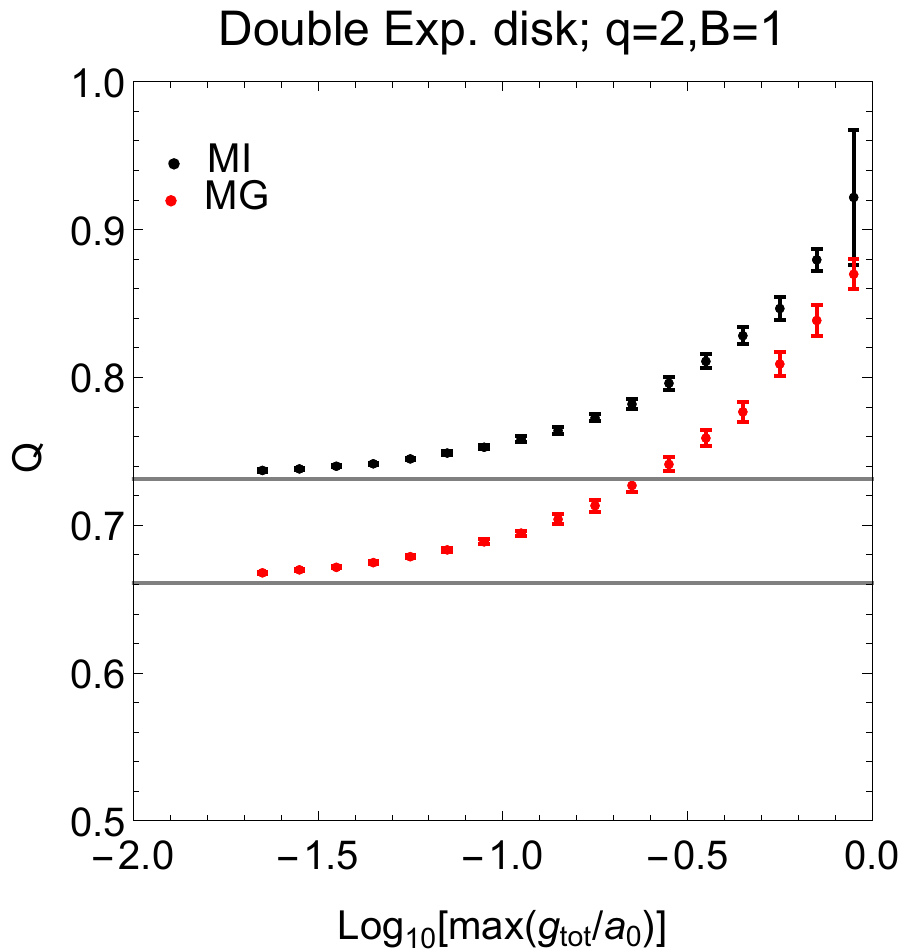}
		\includegraphics[width=0.24\textwidth]{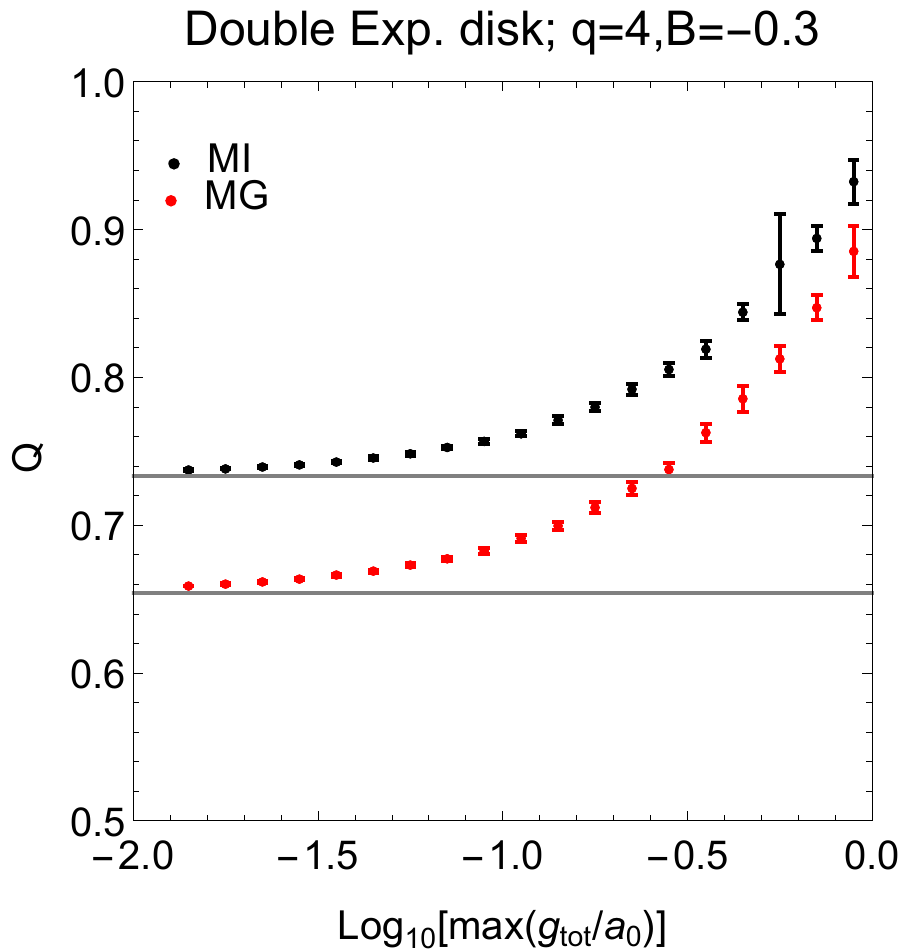}
		\includegraphics[width=0.24\textwidth]{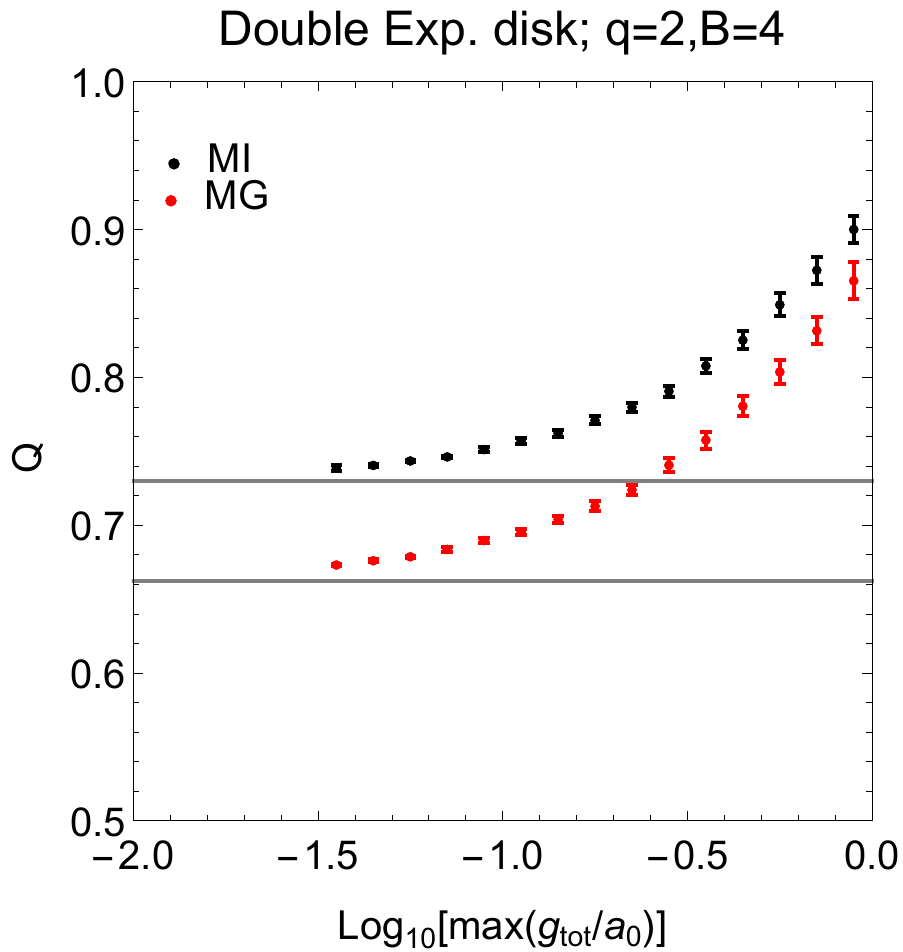}
		\includegraphics[width=0.24\textwidth]{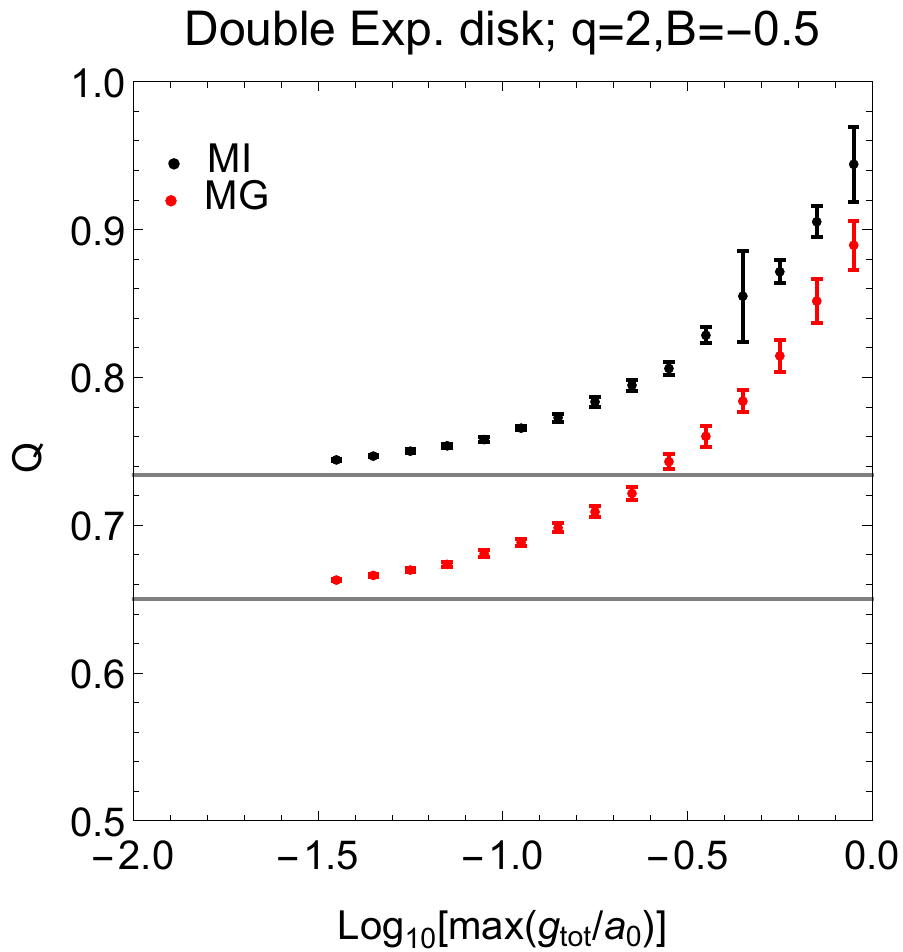}
		\includegraphics[width=0.24\textwidth]{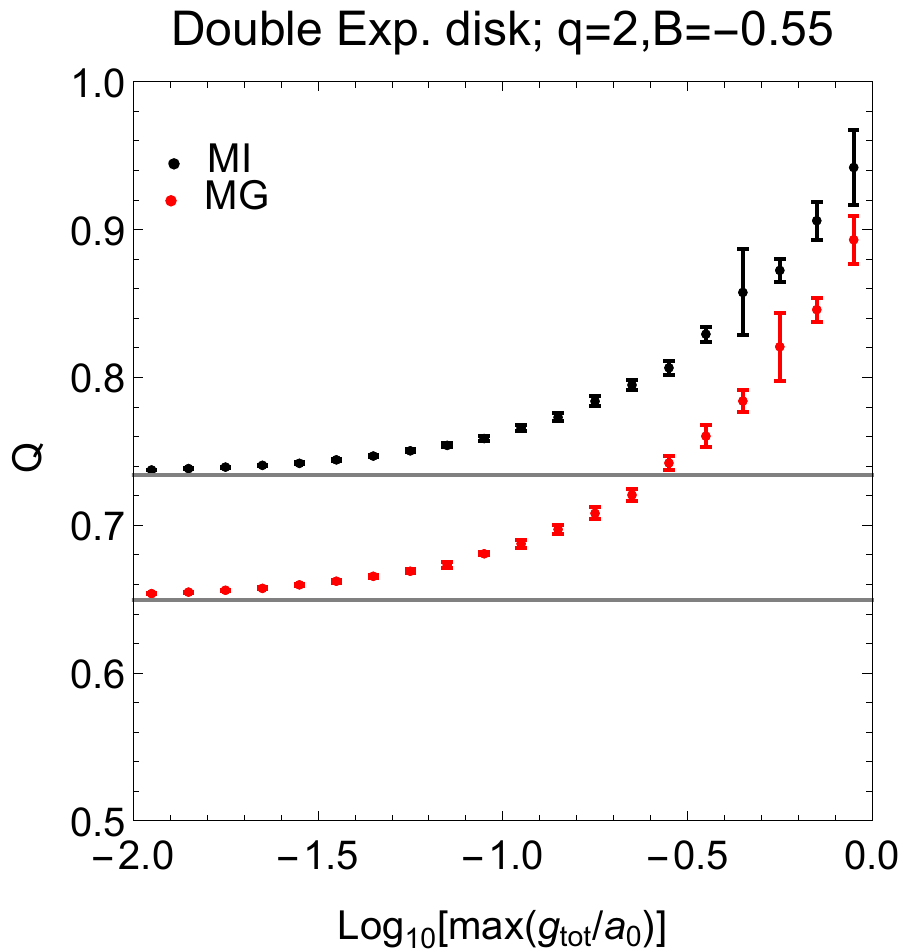}
		\includegraphics[width=0.24\textwidth]{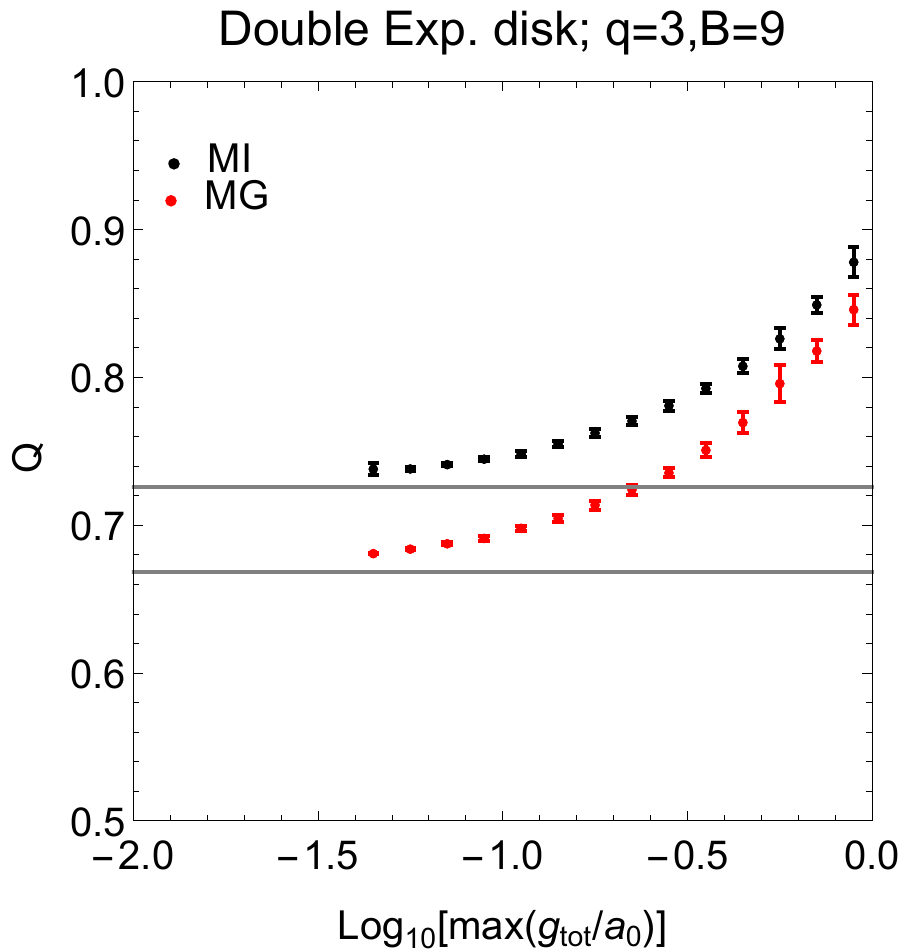}
		\includegraphics[width=0.24\textwidth]{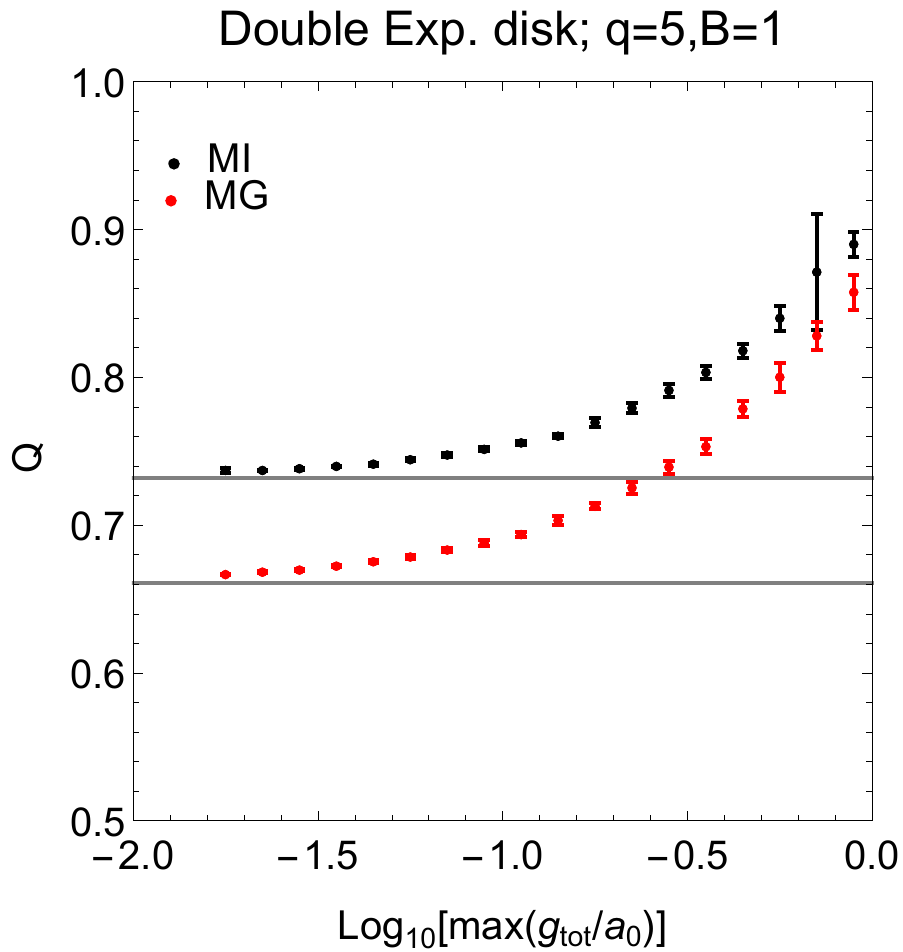}
		\includegraphics[width=0.24\textwidth]{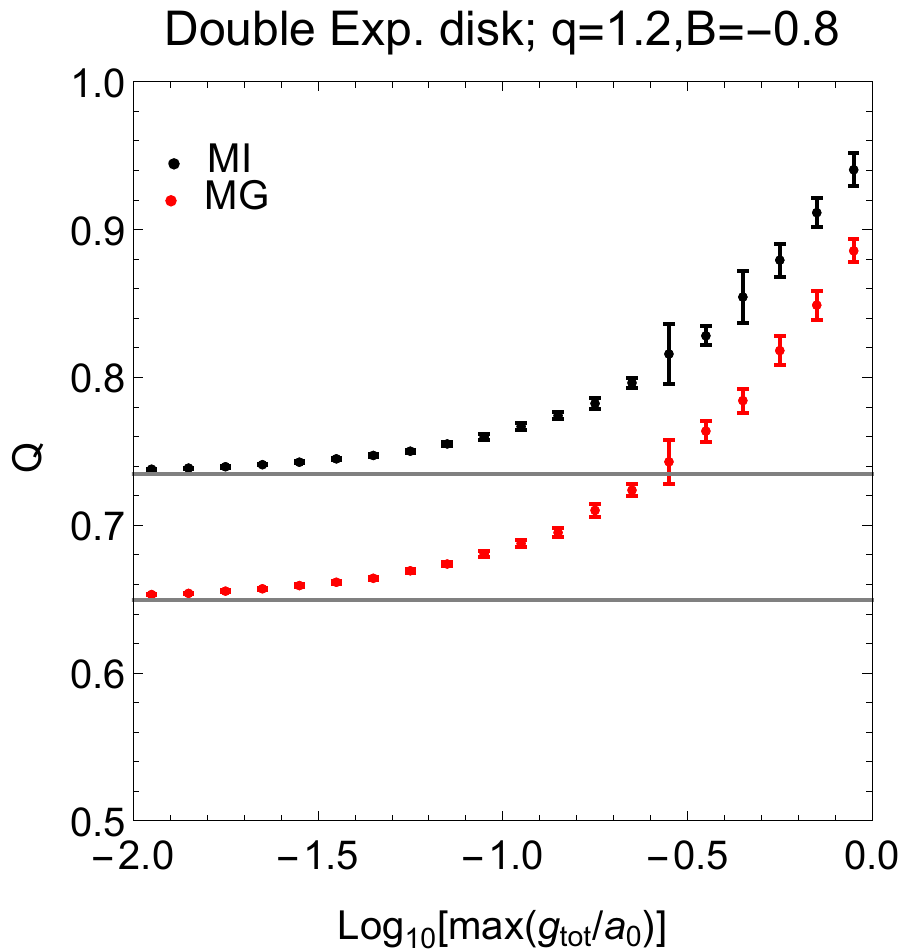}
		\includegraphics[width=0.24\textwidth]{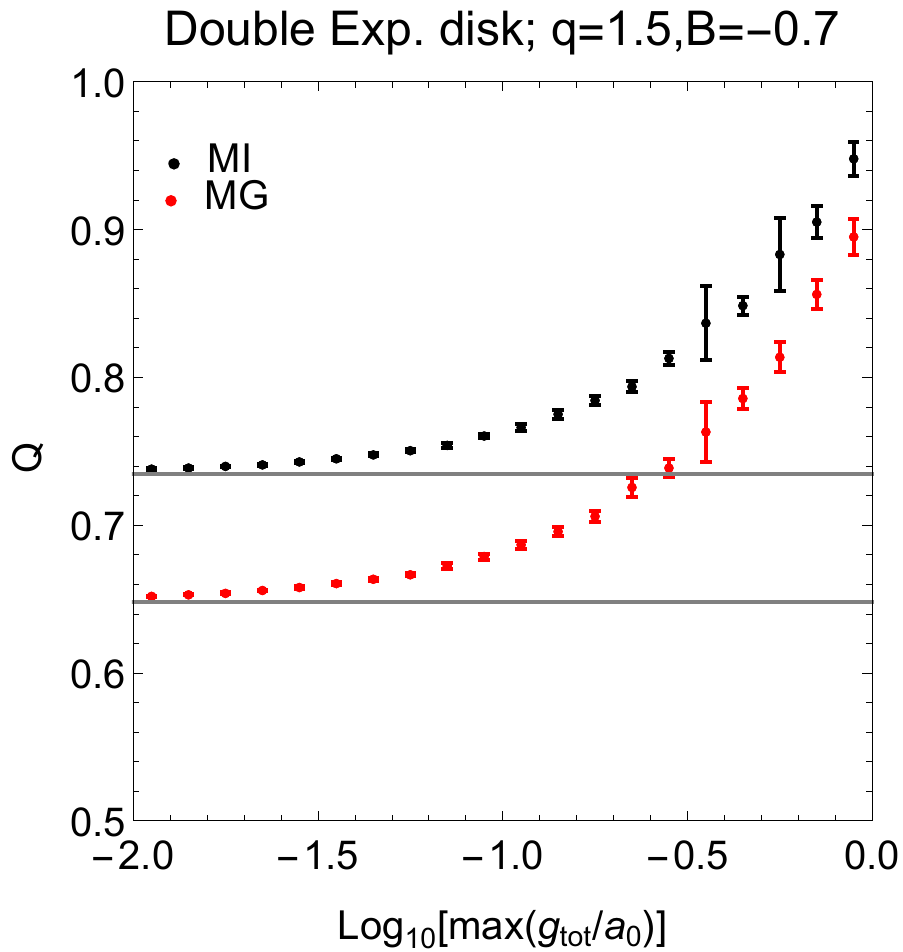}
		\includegraphics[width=0.24\textwidth]{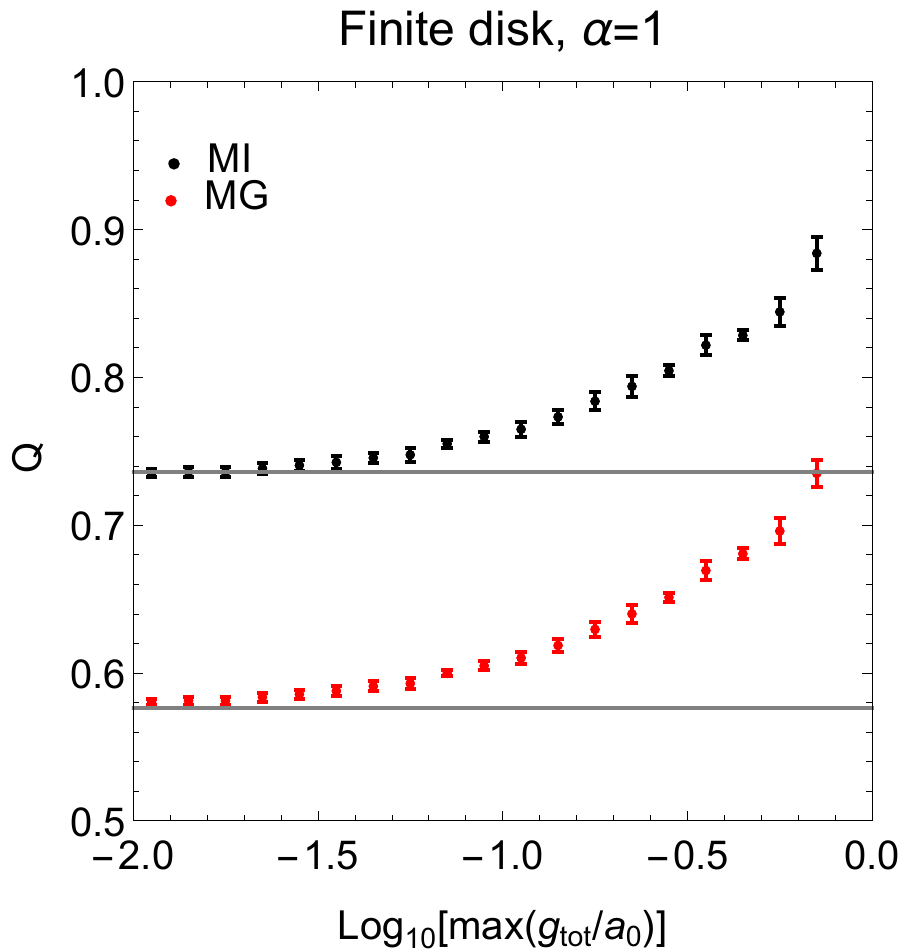}
		\includegraphics[width=0.24\textwidth]{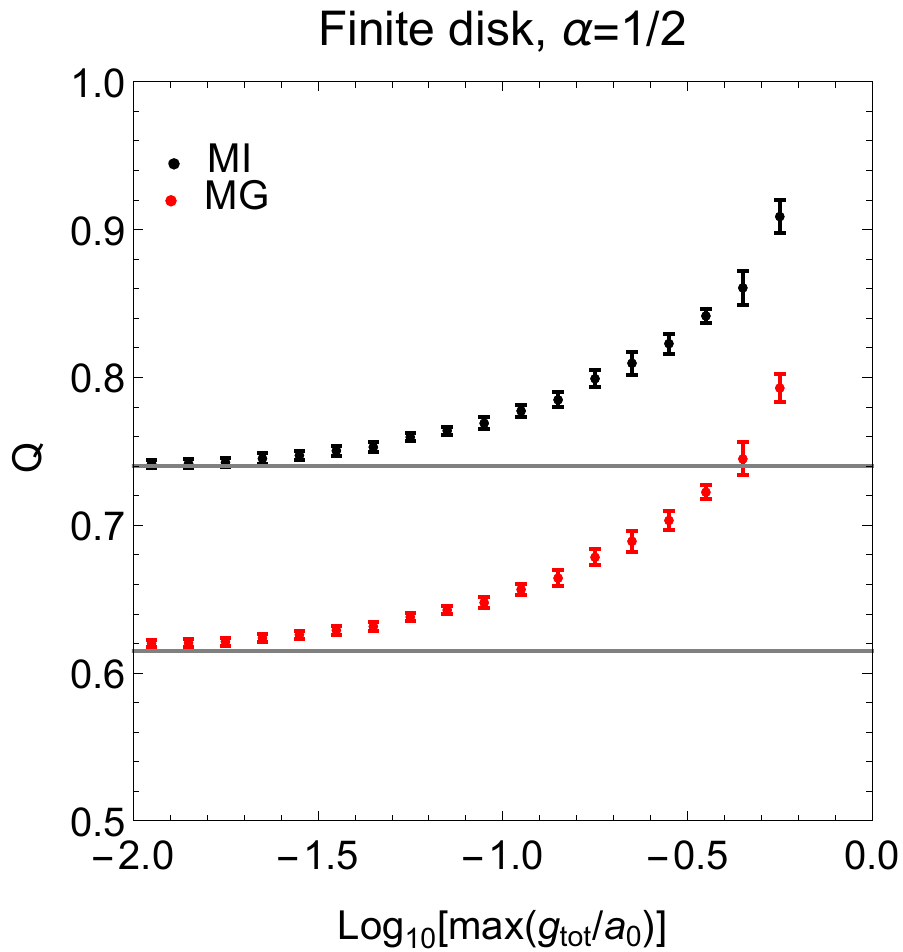}
		\includegraphics[width=0.24\textwidth]{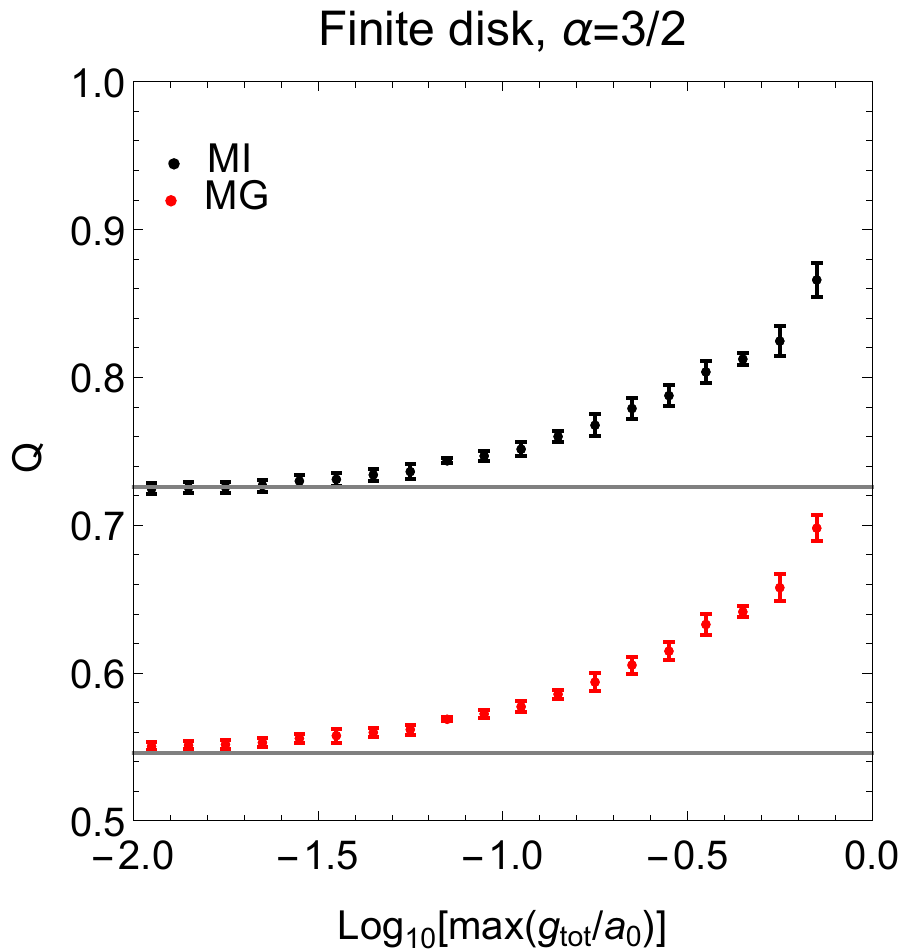}
		\caption{$Q$ calculated from MOND modified inertia and MOND modified gravity in the \citet{Brada:1994pk} approximation from the mock data binned into acceleration bins. Each panel represents a different mass distribution (denoted on the panel). See Eqs. \eqref{eq13} and \eqref{eq14} for details.}
		\label{fig:ti1}
	\end{figure*}

	Here we consider the approach to the DML for the different mass distributions considered in \citet{Milgrom:2012rk};
	\begin{equation}
	\begin{split}
	&\text{Kuzmin disk:}\\
	&\Sigma_d(r)=\Sigma_{0}\bigg(1+\frac{r^2}{r_d^2}\bigg)^{-\frac{3}{2}},\\
	&\text{Exponential disk:}\\
	&\Sigma_d(r)=\Sigma_{0}e^{-\frac{r}{r_d}},\\
	&\text{Double exponential disk:}\\
	&\Sigma_d(r)=\Sigma_{0}\bigg[e^{-\frac{r}{r_d}}+Be^{-q\frac{r}{r_d}}\bigg],\\
	&\text{Finite galaxy disk:}\\&\Sigma_d(r)=\frac{(2\alpha+1)m_d}{2\pi r_d^2}\sqrt{1-\bigg(\frac{r}{r_d}\bigg)^2}\bigg(\frac{r}{r_d}\bigg)^{2(\alpha-1)}F_{21},\qquad\qquad\quad\\
	\end{split}
	\label{eq13}
	\end{equation}
	with
	\begin{equation}
	F_{21}\equiv \,_2F_1\bigg[1-\alpha,\frac{1}{2},\frac{3}{2},1-\bigg(\frac{r}{r_d}\bigg)^{-2}\bigg]
	\label{eq14}
	\end{equation}
	being hypergeometric functions. For the MG calculation, we adopt the approximation from \citet{Brada:1994pk}. The results are shown in Figure \ref{fig:ti1}.

\section{Galaxy set}
\label{app:rot}
\begin{figure*}[ht]
	\centering
	\captionsetup{width=1\textwidth}
	\includegraphics[width=0.24\textwidth]{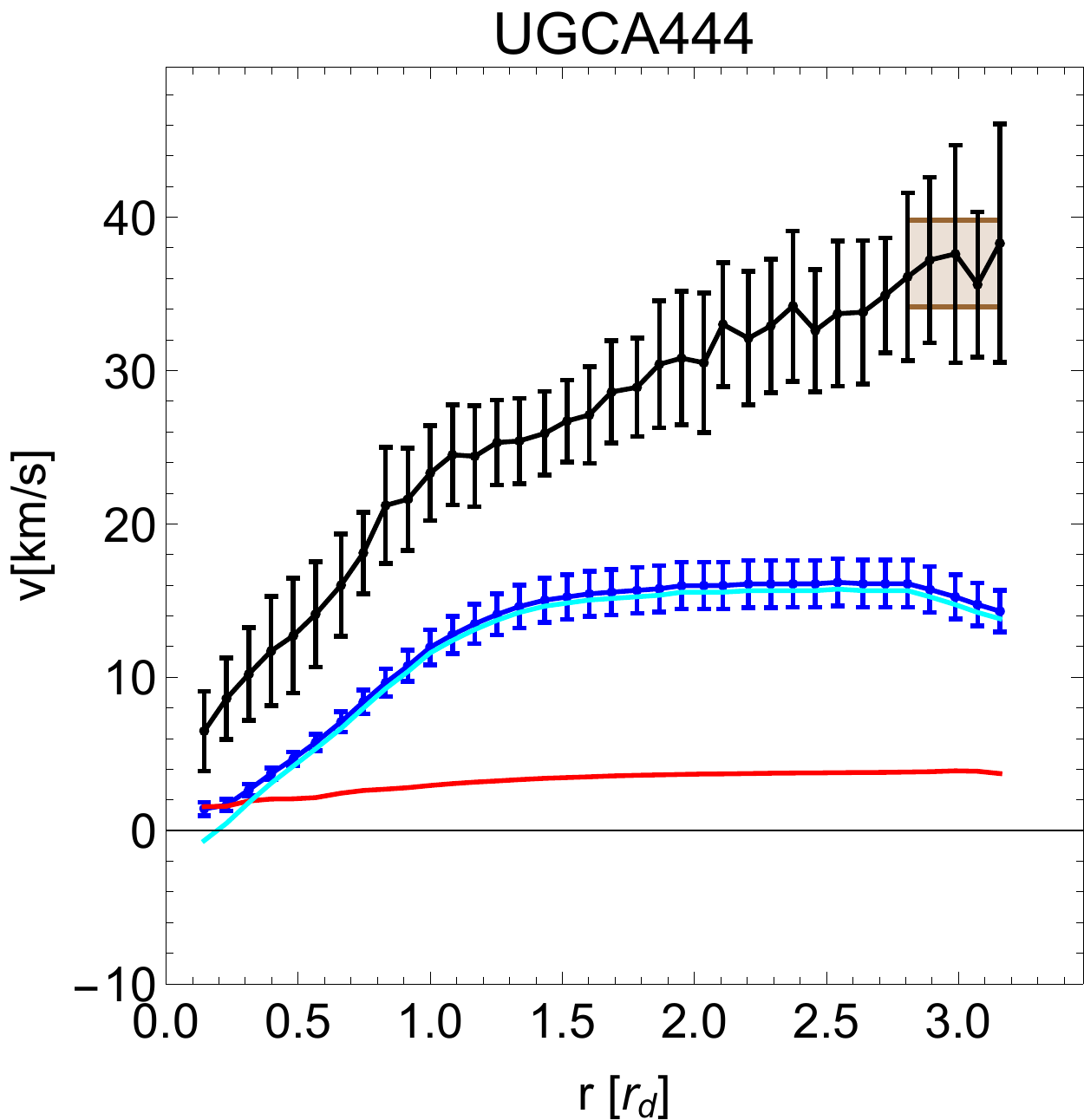}
	\includegraphics[width=0.24\textwidth]{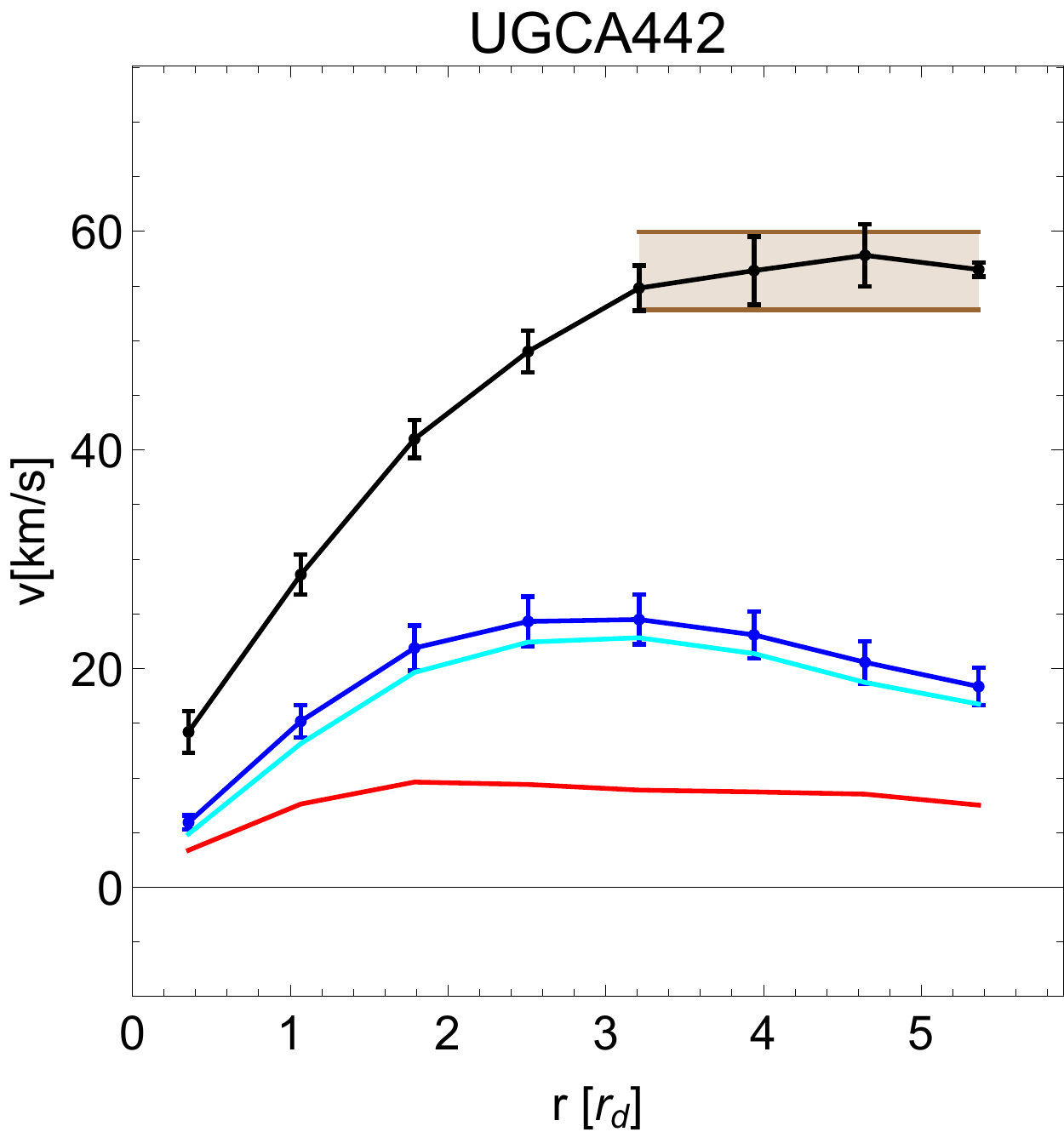}
	\includegraphics[width=0.24\textwidth]{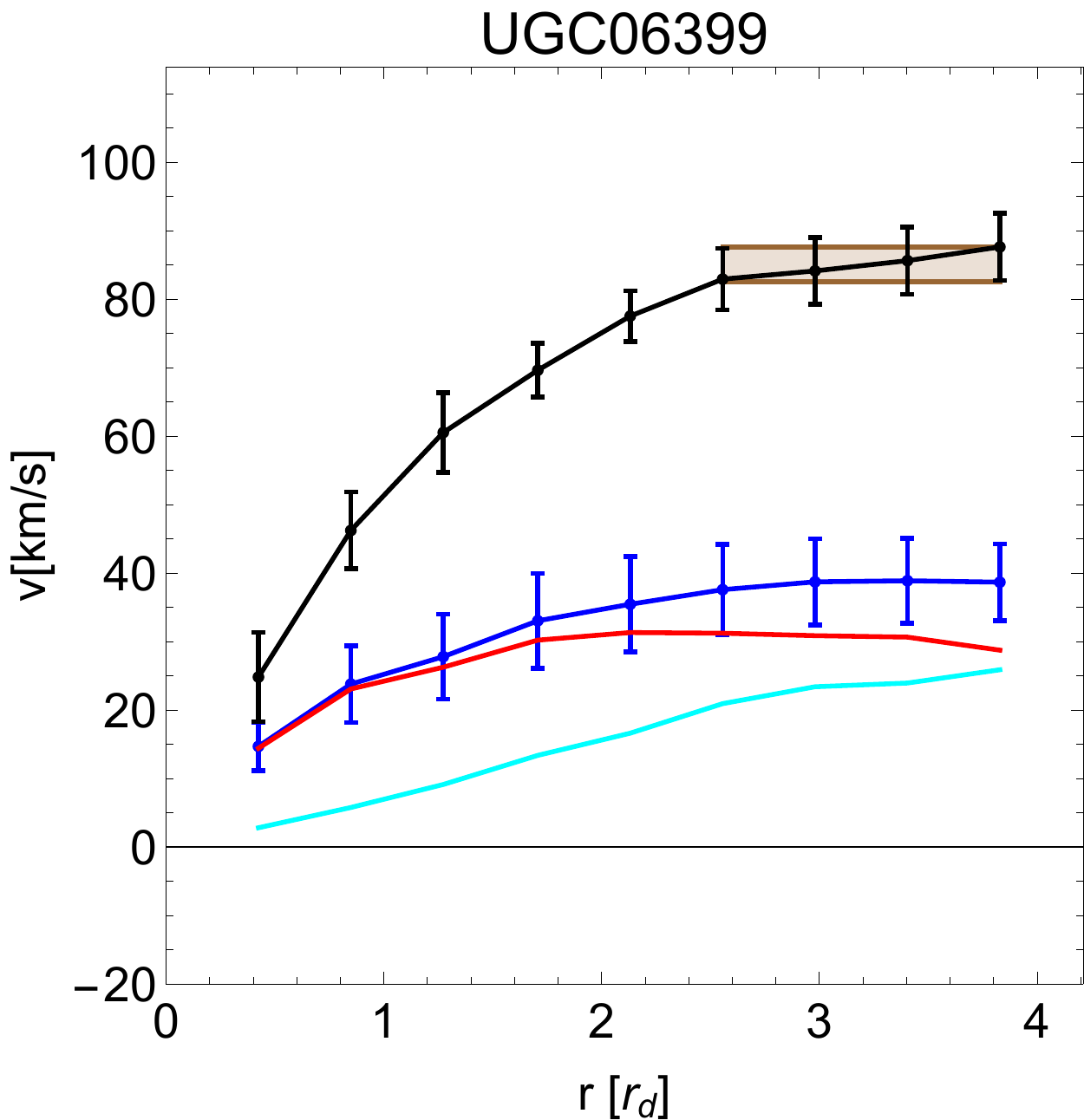}
	\includegraphics[width=0.24\textwidth]{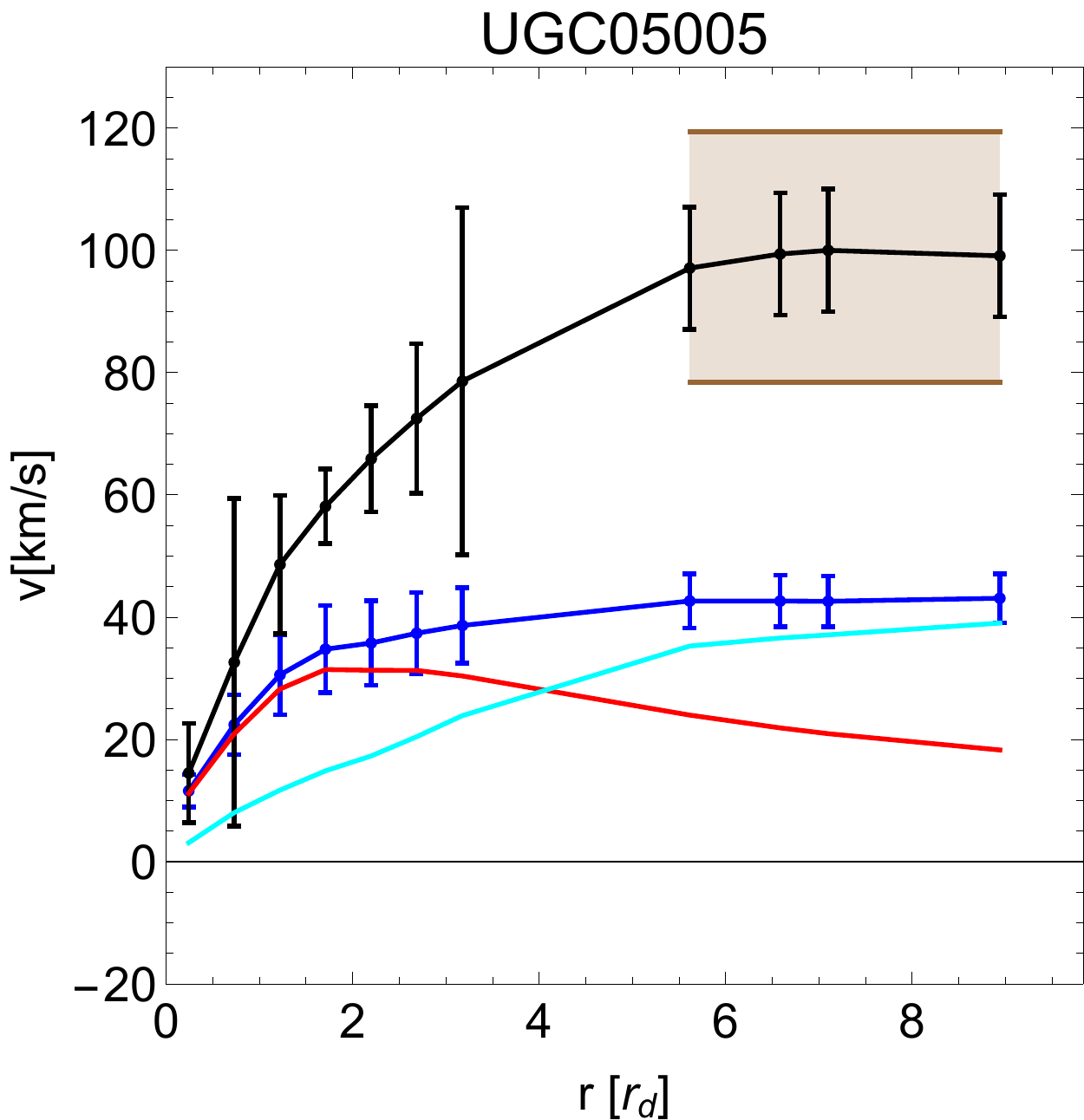}
	\includegraphics[width=0.24\textwidth]{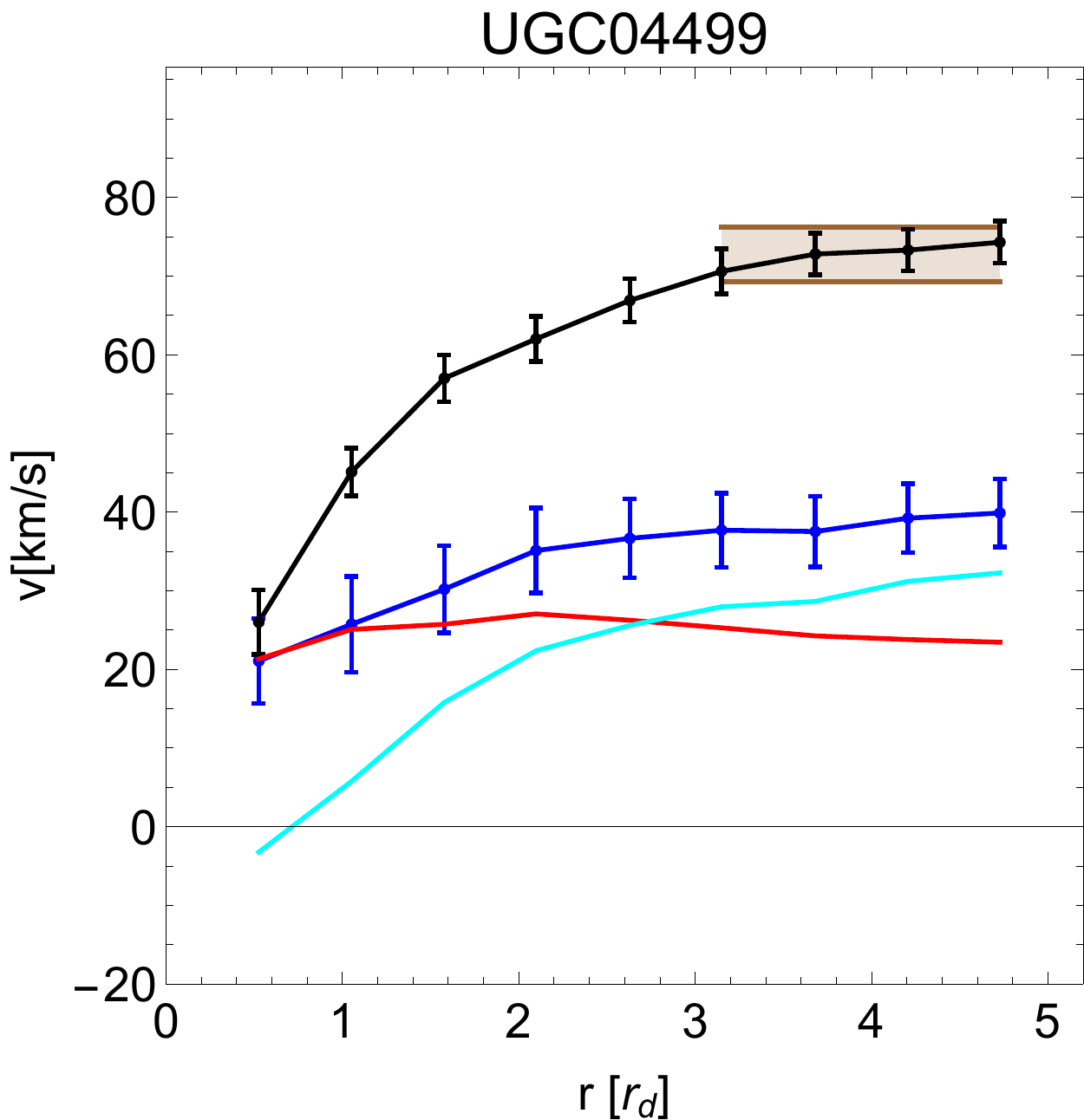}
	\includegraphics[width=0.24\textwidth]{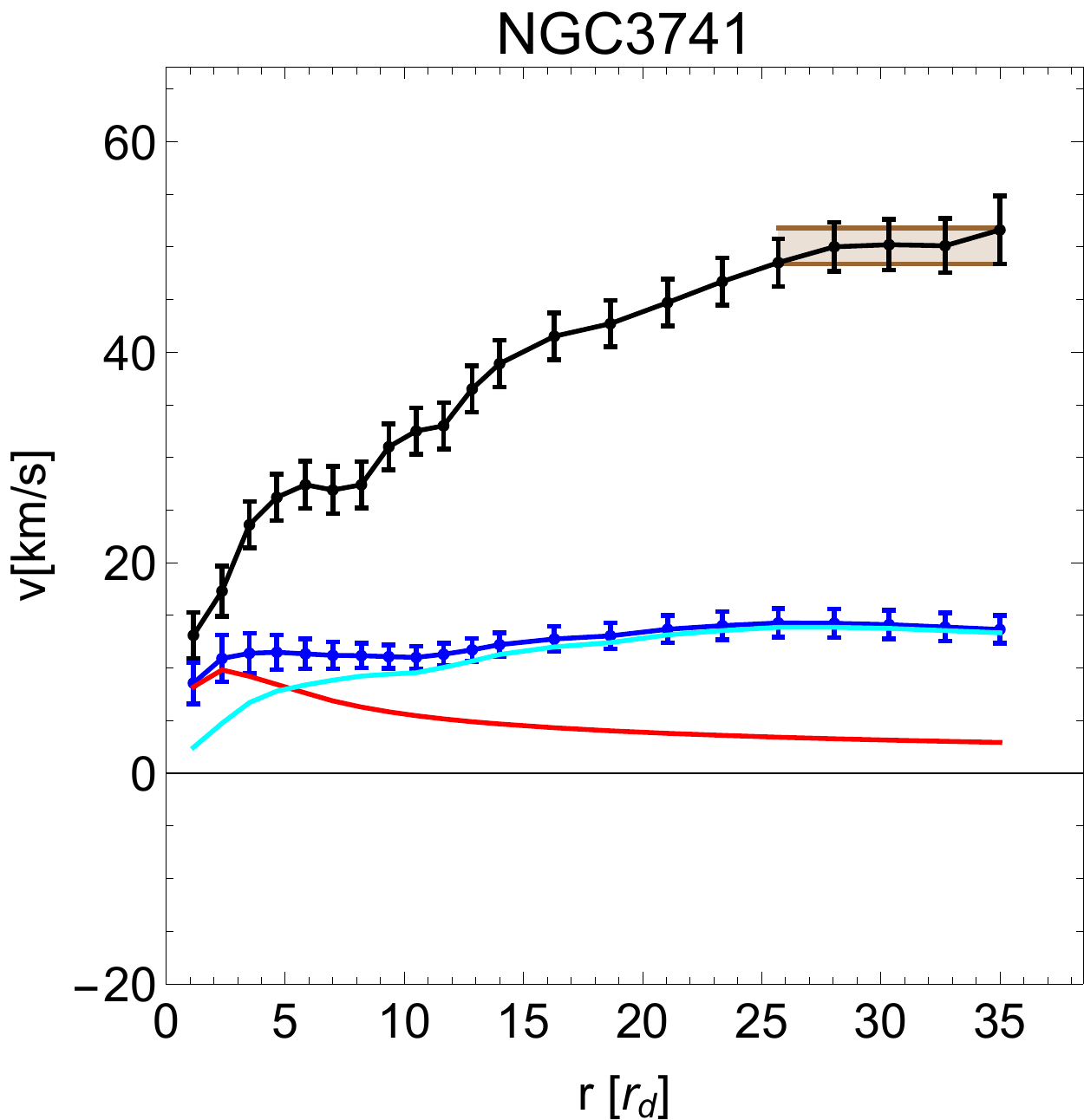}
	\includegraphics[width=0.24\textwidth]{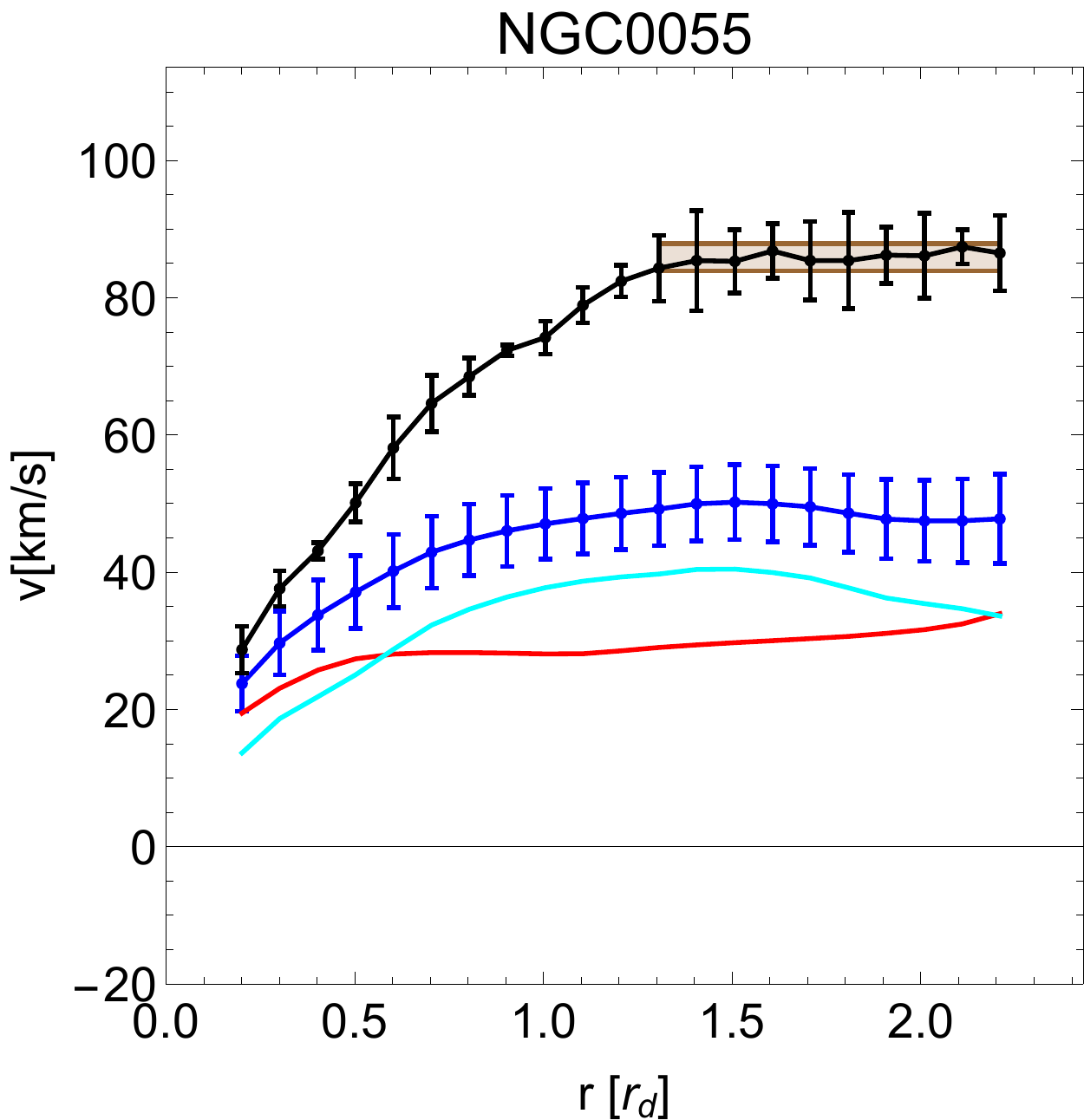}
	\includegraphics[width=0.24\textwidth]{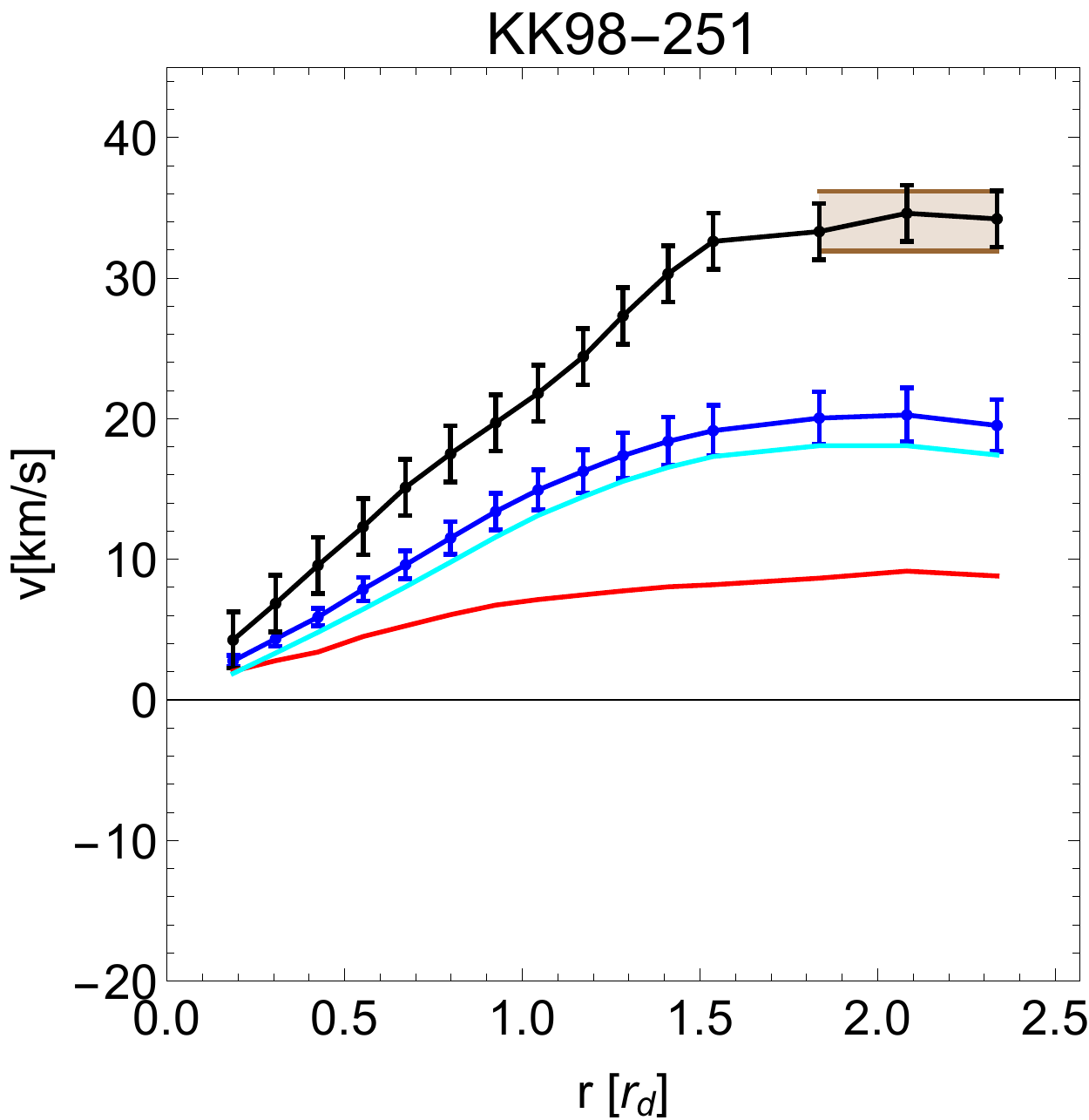}
	\includegraphics[width=0.24\textwidth]{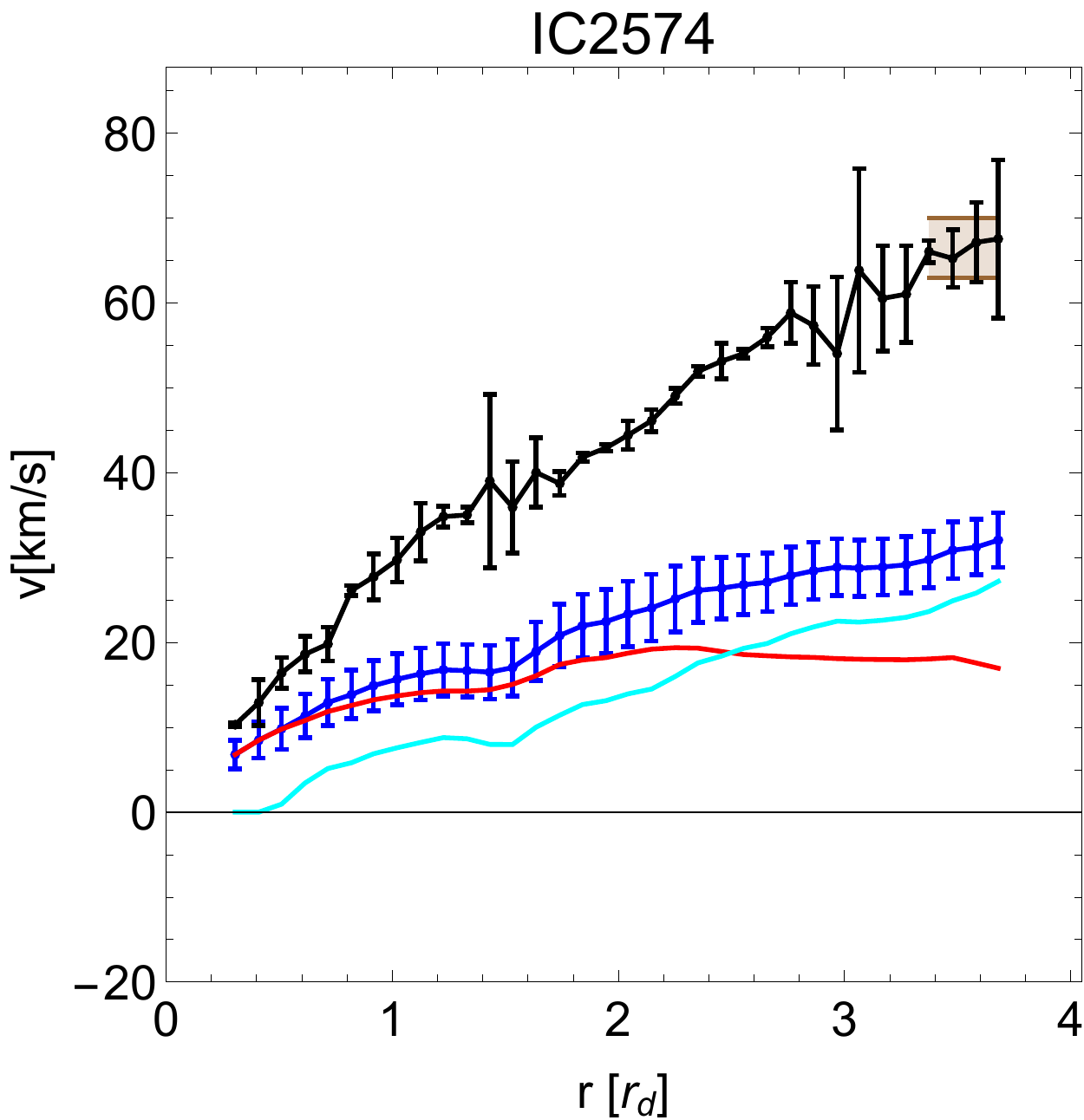}
	\includegraphics[width=0.24\textwidth]{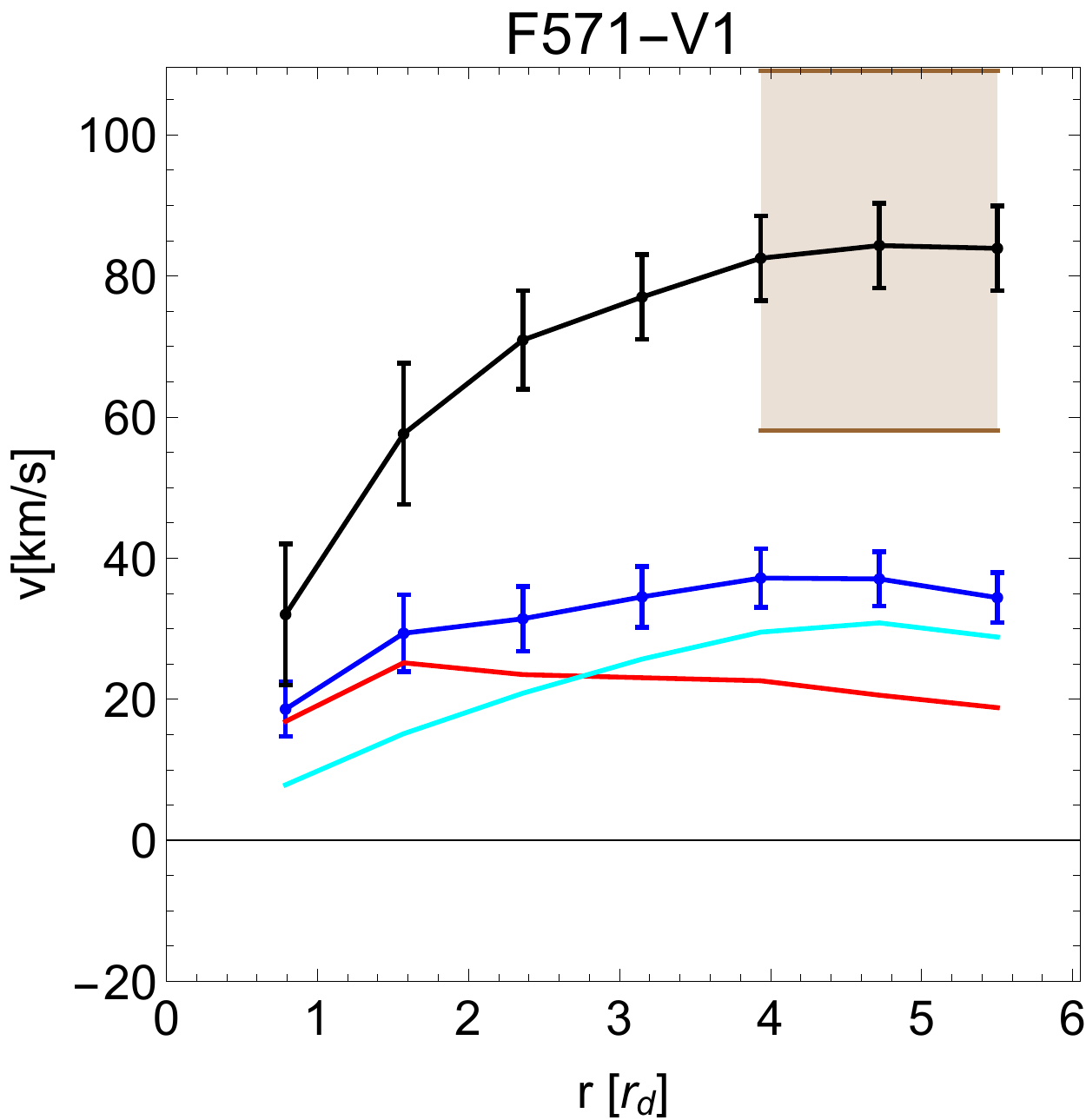}
	\includegraphics[width=0.24\textwidth]{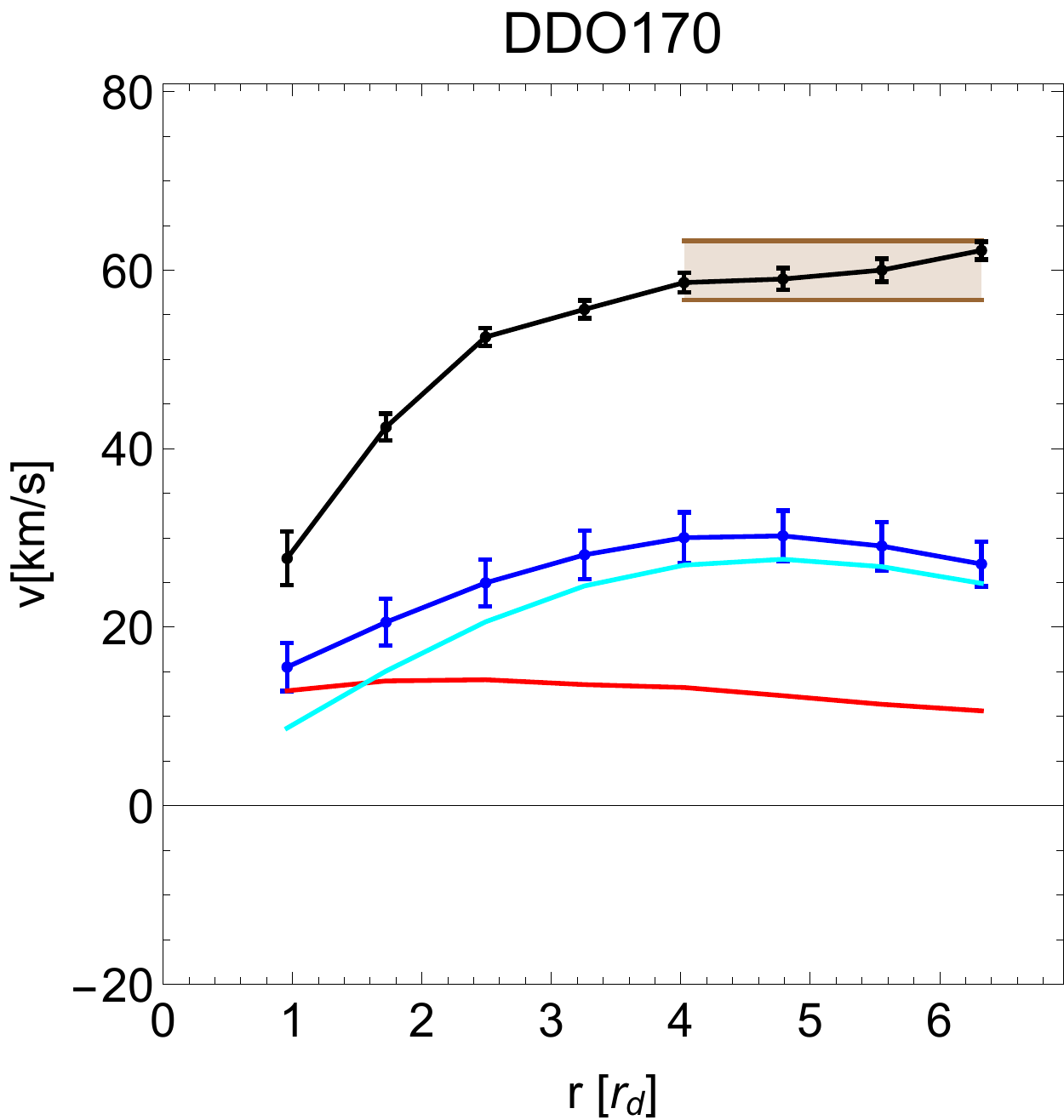}
	\includegraphics[width=0.24\textwidth]{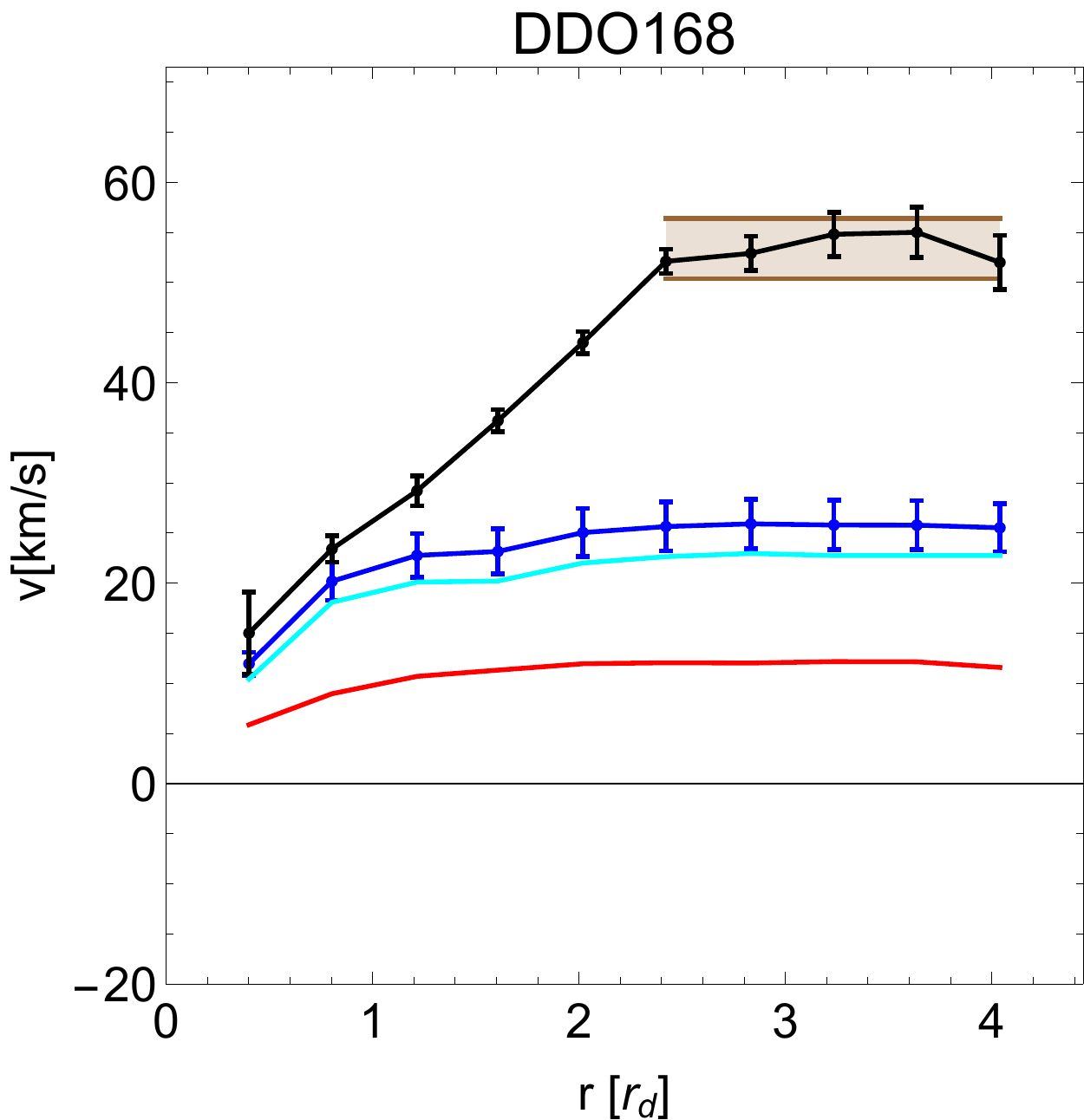}
	\includegraphics[width=0.24\textwidth]{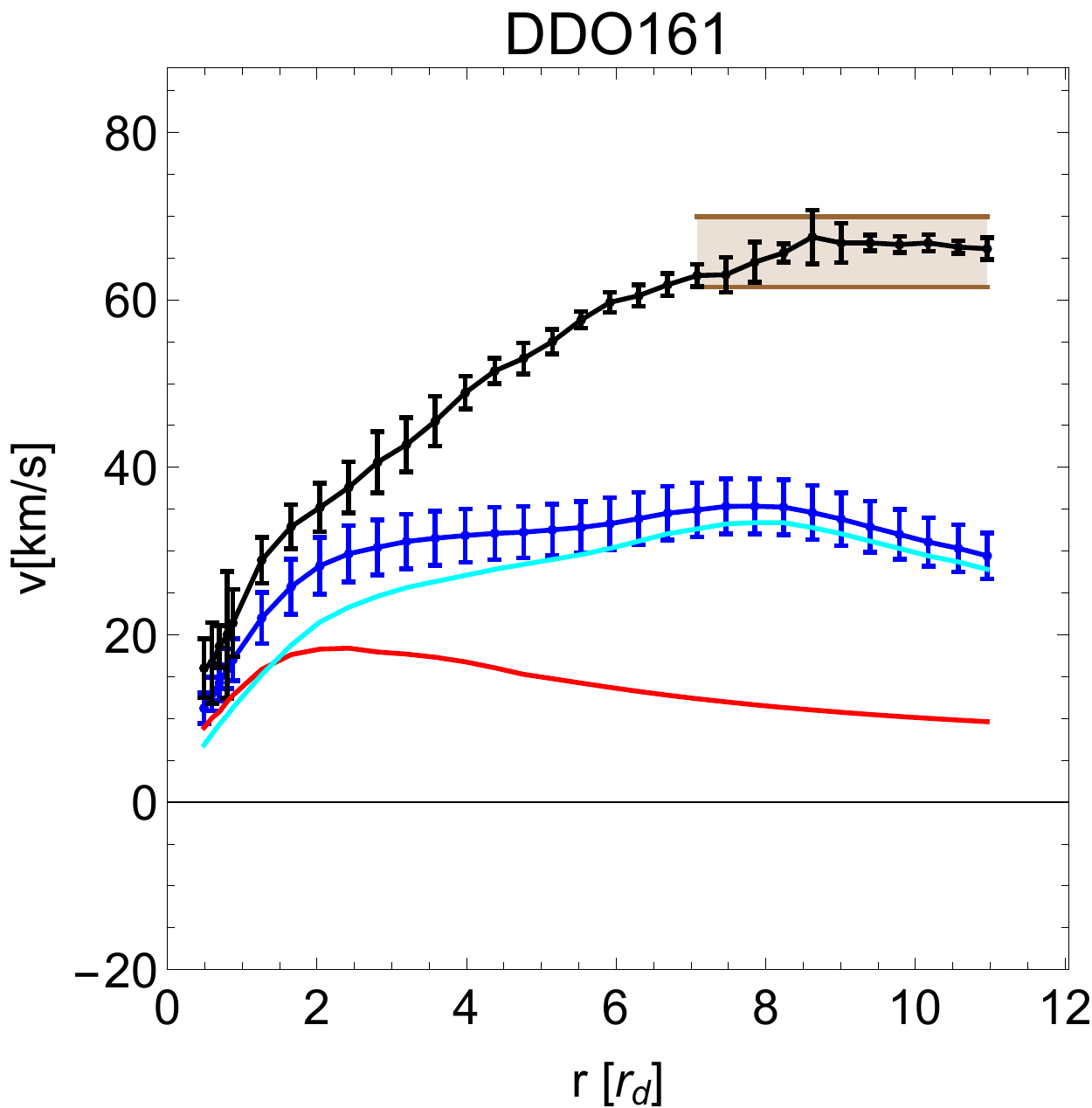}
	\includegraphics[width=0.24\textwidth]{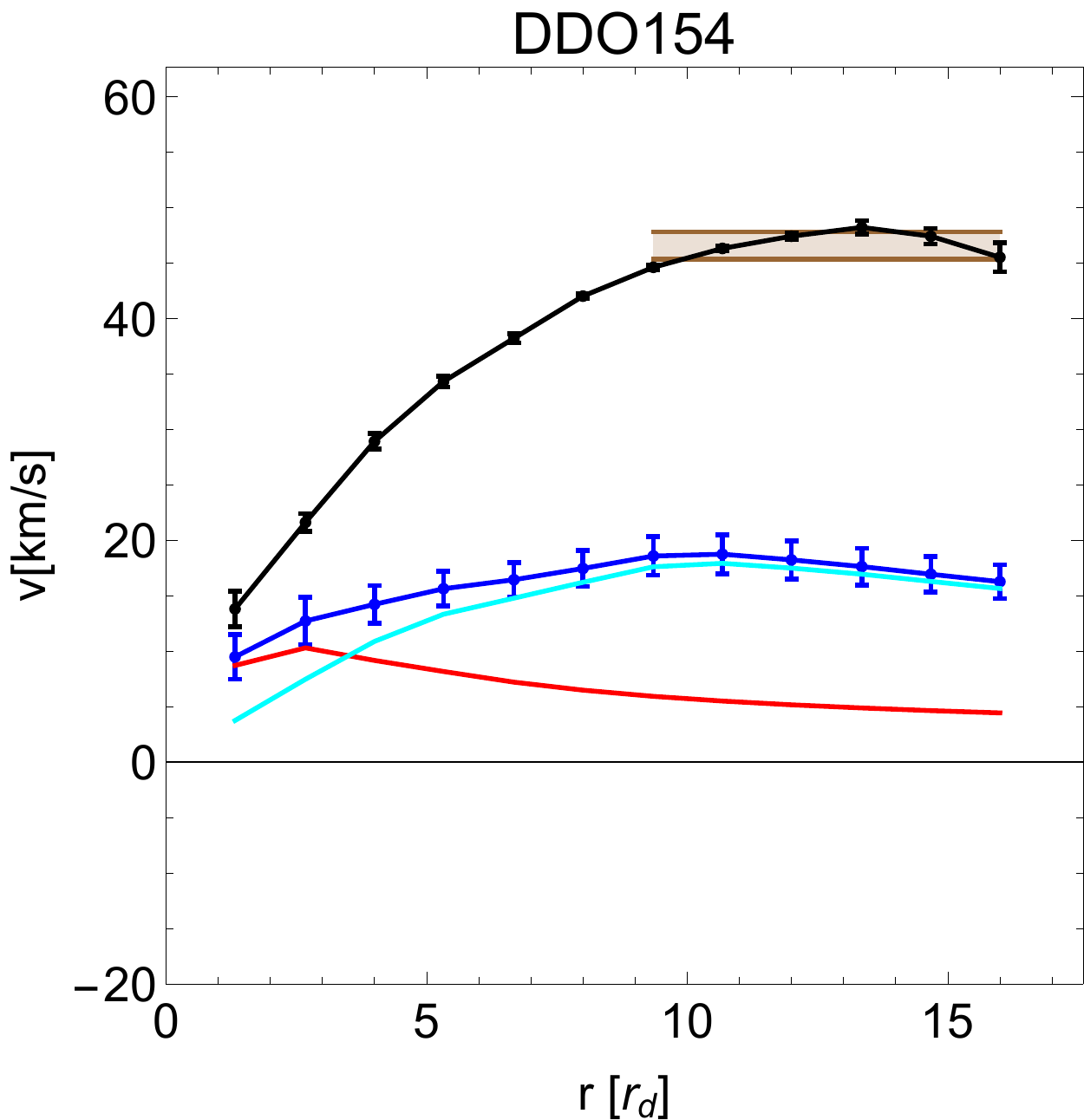}
	\includegraphics[width=0.24\textwidth]{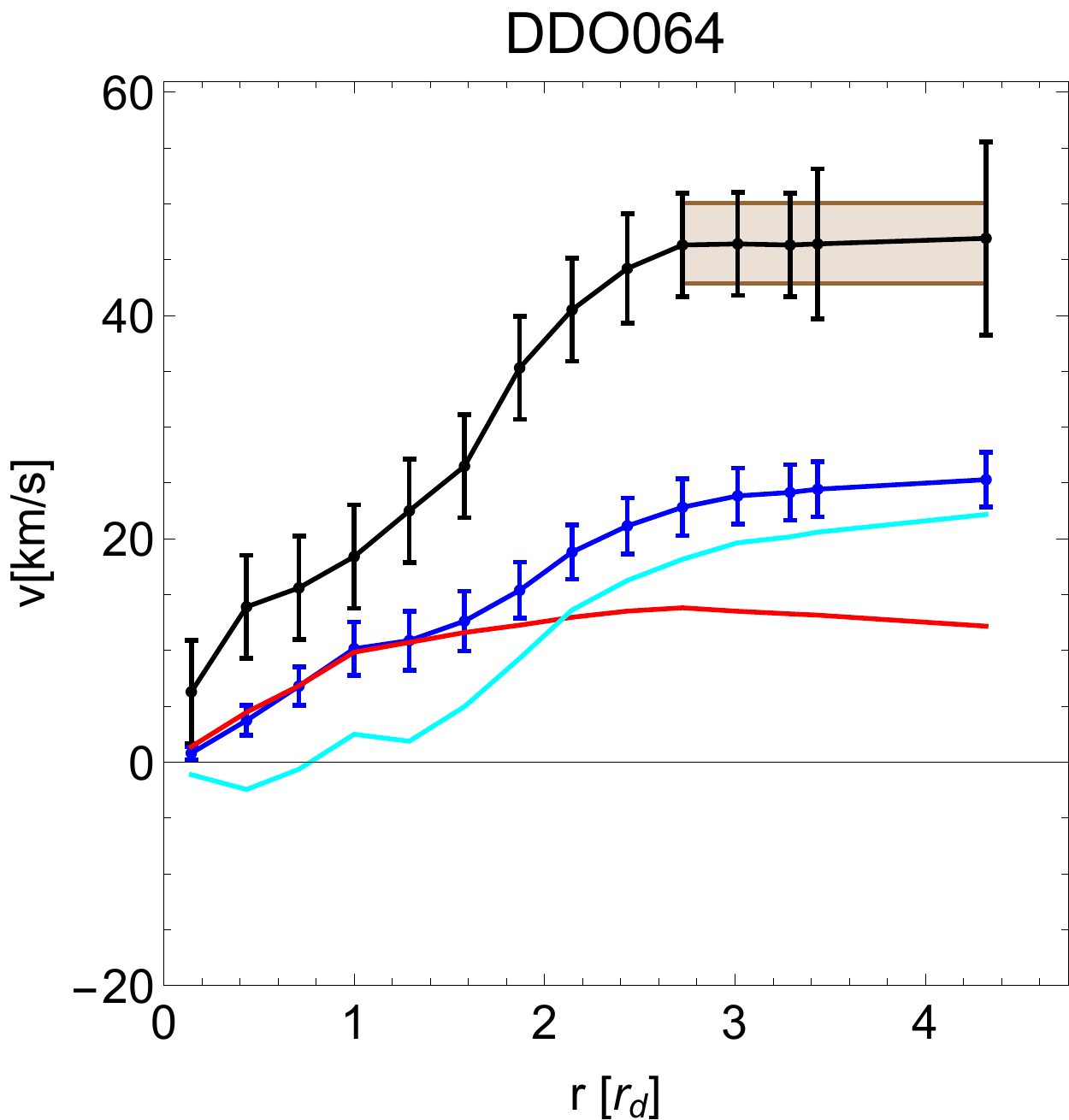}
	\caption{Rotation curves of galaxies in the SPARC database after imposing $\max[g_{obs}]\leq 0.4a_0$, $\frac{\max[r]}{r^d}\geq 3$ and spacing $\leq 1.3 r_d$ as well as requiring that there be a gas profile and a well-defined approximation of $v_\infty$. Black is $v_{obs}$, blue is $v_{bar}$, cyan is $v_g$, red is $v_d$ and brown denotes the $1\sigma$ region for $v_f$. The brown band illustrates the points used to compute $v_f$.}
	\label{fig:5f}
\end{figure*}

Here we show the mass models for the $15$ galaxies (see figure \ref{fig:5f}) that satisfy our quality criteria: $\max(g_{obs})< 0.4a_0$, $\max(r)>3 r_d$, and rotation-curve spacing smaller than $1.3r_d$. The properties of these $15$ galaxies are given in Table 4\ref{table:4}.

\begin{table*}[ht]
	\caption{Galaxy set} % title of Table
	\label{table:4} % is used to refer this table in the text
	\centering % used for centering table
	\begin{tabular}{c c c c c c } % centered columns (4 columns)
		\hline\hline % inserts double horizontal lines
		Galaxy & $Q^{(m)}$ & $Q^{(c)}$ & $\max(g_{obs})$& $\max(r)$ & Spacing\\
		 &  &  & $[a_0]$& $[r_d]$ & $[r_d]$\\
		\hline % inserts single horizontal line
		UGCA444 &$ 0.50 \pm 0.04$ & $ 0.51 \pm 0.05$ & $0.18$& $4.95$ & $0.14$\\
		UGCA442 &$ 0.55 \pm 0.01$ & $ 0.47 \pm 0.02$ & $0.22$& $7.82$ & $0.98$\\
		UGC06399& $ 0.70 \pm 0.04$& $0.69\pm 0.04$ & $0.38$ & $3.26$ & $0.36$\\
		UGC05005&$ 0.64 \pm 0.15$ & $ 0.65\pm 0.15 $ & $0.17$& $3.66$ & $0.33$\\
		UGC04499&$ 0.74 \pm 0.03$ & $ 0.72\pm 0.03 $ & $0.33$& $3.18$ & $0.35$\\
		NGC3714&$ 0.59 \pm 0.03$ & $0.59\pm 0.03$ & $0.22$& $4.04$ & $0.19$\\
		NGC0055&$ 0.69 \pm 0.02$ & $0.68\pm 0.02$ & $0.26$& $4.53$ & $0.22$\\
		KK98-251&$ 0.57 \pm 0.05$ & $0.60\pm 0.05$& $0.14$& $3.81$ & $0.25$\\						
		IC2574& $ 0.48 \pm 0.02$& $0.52\pm 0.03$  & $0.13$& $3.23$ & $0.10$\\
		F571-V1&$ 0.71 \pm 0.06$ & $0.66\pm 0.07$ & $0.23$& $5.27$&$0.75$\\				
		DDO170&$ 0.75 \pm 0.02$ & $0.74\pm 0.02$  & $0.15$& $5.40$&$0.68$\\
		DDO168&$ 0.61 \pm 0.02$ & $0.61\pm 0.02$  & $0.30$& $3.44$ &$0.34$\\
		DDO161&$ 0.57 \pm 0.02$ & $0.58\pm 0.02$  & $0.15$&$6.09$ & $0.20$\\
		DDO154&$ 0.72 \pm 0.01$ & $0.74\pm 0.02$  & $0.16$&$3.38$ & $0.28$\\		
		DDO064&$ 0.66 \pm 0.08$ & $0.66\pm 0.08$  & $0.31$& $3.08$ &$0.22$\\
		\hline %inserts single line
	\end{tabular}
\end{table*}
\clearpage

\bibliographystyle{aa}
\bibliography{refs} 

\end{document}